\documentclass[aps,prx,twocolumn,superscriptaddress,showpacs,amsmath,amssymb]{revtex4-1}
\usepackage{amssymb}
\usepackage{graphicx}
\usepackage{bm}
\usepackage{amsmath}
\usepackage{color} 

\newcommand{\beq}{\begin{equation}}
\newcommand{\eeq}{\end{equation}}
\newcommand{\beqn}{\begin{eqnarray}}
\newcommand{\eeqn}{\end{eqnarray}}

\renewcommand{\vec}[1]{\mbox{\boldmath$#1$}}
\usepackage[unicode=true,
 bookmarks=true,bookmarksnumbered=false,bookmarksopen=false,
 breaklinks=false,pdfborder={0 0 1},backref=false,colorlinks=true]
 {hyperref}
\hypersetup{
 linkcolor=magenta, urlcolor=blue, citecolor=blue, pdfstartview={FitH}, hyperfootnotes=false, unicode=true}

\begin{document}

\title{Majorana spintronics} 
\author{Xin Liu}
\affiliation{School of Physics and Wuhan National High Magnetic Field Center, Huazhong University of
Science and Technology, Wuhan, Hubei 430074, China}
\affiliation{Condensed Matter Theory Center and Joint Quantum
Institute, Department of Physics, University of Maryland,
College Park, MD 20742-4111, USA}
\author{Xiaopeng Li}
\affiliation{Condensed Matter Theory Center and Joint Quantum
Institute, Department of Physics, University of Maryland,
College Park, MD 20742-4111, USA}
\author{Dong-Ling Deng}
\affiliation{Condensed Matter Theory Center and Joint Quantum
Institute, Department of Physics, University of Maryland,
College Park, MD 20742-4111, USA}
\author{Xiong-Jun Liu}
\affiliation{ International Center for Quantum Materials and School of Physics, Peking University, Beijing 100871, China}
\affiliation{Collaborative Innovation Center of Quantum Matter, Beijing 100871, China}
\author{S. Das Sarma}
\affiliation{Condensed Matter Theory Center and Joint Quantum
Institute, Department of Physics, University of Maryland,
College Park, MD 20742-4111, USA}
\date{\today}

\begin{abstract}
We propose a systematic magnetic-flux-free approach to detect, manipulate and braid Majorana fermions  in a semiconductor nanowire-based topological Josephson junction by utilizing the Majorana spin degree of freedom. We find an intrinsic $\pi$-phase difference between spin-triplet pairings enforced by the Majorana zeros modes (MZMs) at the two ends of a one-dimensional spinful topological superconductor. This $\pi$-phase is identified to be a spin-dependent superconducting phase, referred to as the spin-phase, which we show to be tunable by controlling spin-orbit coupling strength via electric gates. This electric controllable spin-phase not only affects the coupling energy between MZMs but also leads to a fractional Josephson effect in the absence of any applied magnetic flux, which enables the efficient topological qubit readout. We thus propose an all-electrically controlled superconductor-semiconductor hybrid circuit to manipulate MZMs and to detect their non-Abelian braiding statistics properties. Our work on spin properties of topological Josephson effects potentially opens up a new thrust for spintronic applications with Majorana-based semiconductor quantum circuits. 
\end{abstract}
\maketitle

\section{Introduction}
Spin is the fundamental electronic quantum degree of freedom in solid state materials. In superconductors, Cooper pairs, as composed of two spin-1/2 particles, can have spin-1 angular momentum, leading to spin-triplet pairings, in contrast to the usual spin-singlet pairing of opposite spins in the simplest $s$-wave superconductors. Recently, it has been shown that Majorana zero modes (MZMs), which may exist as stable localized zero energy mid-gap excitations in topological superconductor interfaces, can only have $s$-wave odd-frequency \cite{TSC:Asano2013} spin-triplet correlations \cite{TSC:He2014,TSC:Liu2015} 
at the boundary of topological superconductors (TSCs) \cite{TSC:Kitaev2001,TSC:Fu2008,TSC:Zhang2008,TSC:Sato2009,TSC:Sau2010,TSC:Lutchyn2010,TSC:Oreg2010,TSC:Sau2010a, TSC:Alicea2010,TSC:Alicea2011,TSC:LiuLobos2013,TSC:Beenakker2013a,TSC:Zhang2013,TSC:Liu2014,TSC:Ebisu2014,TSC:Sun2014,TSC:Hui2015,TSC:Brydon2015}. Since the $s$-wave spin-triplet pairing is insensitive to non-magnetic impurity scattering, spin-triplet pairs can be stabilized at the interface of a topological-superconductor/normal-metal (TSC/NM) hybrid system. Consequently, MZMs can assist the injection of pure stable spin-triplet Cooper pairs into the normal-metal region of a TSC/NM/TSC junction. Thus, a topological Josephson junction (JJ) is indeed a spin-triplet JJ which makes it possible to utilize the Majorana spin degree of freedom to detect and manipulate MZMs, and demonstrate their non-Abelian braiding statistics. 

\begin{figure}[htp]
\centering
\begin{tabular}{l}
\includegraphics[width=0.9\columnwidth]{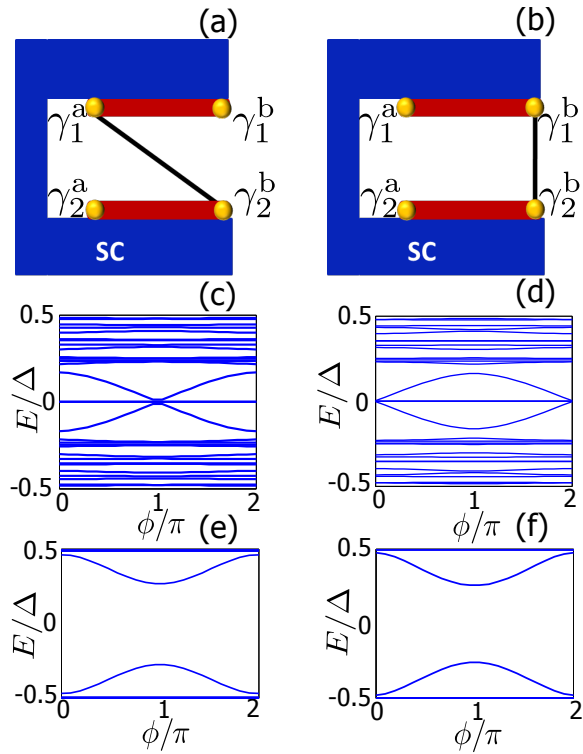}
\end{tabular}
\caption{(a) and (b) The Josephson junctions with N-shape and U-shape geometries. The red wires are the semiconductor wires attached to the $s$-wave superconductor. The yellow balls indicate the locations of MZMs. The black wires represent the normal metal. Panels (c) and (d) show the Andreev levels from the numerical calculations of the eigenenergy of the Josephson junctions in (a) and (b) with $t_{\rm so}=\Delta$, $M=8\Delta$, $\mu_s=-2\Delta$, $ t_s=t_{\rm N}=10\Delta$, $\mu_N=20\Delta.$ 
Panels (e) and (f) show the Andreev levels of the Josephson junctions in (a) and (b) using the same set of parameters except changing $M$ to be $\Delta/2$.}
\label{D-current-phase}
\end{figure}

In this work, we theoretically study the spin dependent current-phase relation and MZM coupling energy of topological JJs. We show that the MZM-induced spin-triplet pairing states \cite{TSC:Liu2015} at the two ends of a realistic one-dimensional (1D) TSC (specifically, the nanowire-superconductor hybrid system of great current interest) have an intrinsic $\pi$-phase difference. We demonstrate that this $\pi$-phase, originating from Majorana enforced spin-triplet pairing, arises neither from a magnetic flux-induced phase, refereed to charge phase since magnetic flux is coupled to electrons\rq{} charge degree of freedom, nor from the Cooper pairs\rq{} orbital (e.g. $p$- or $d$-wave) effect, and can produce a fractional Josephson $0$-junction and $\pi$-junction, which exhibit a Josephson phase of $0$ and $\pi$ in its ground state in the absence of any applied magnetic flux respectively, in the N-shape and U-shape nanowires  [Fig.~\ref{D-current-phase}(a,b)]. From this result, we unambiguously establish the presence of a spin-state dependent phase, referred to as {\it spin-phase}, in spin-triplet pairings in this system. The spin-phase difference across the topological JJ can be continuously tuned by spin-orbit coupling (SOC) in the normal part (black wire) of the JJ through a gate voltage [Fig.~\ref{H-JJ}(a)] so as to turn on and off the MZM coupling energy for both time-reversal invariant and time-reversal broken topological JJs leading to experimentally-testable $4\pi$ periodic Josephson current-phase relations in both charge phase and spin-phase. In particular, the SOC tunable spin-phase can drive the time-reversal symmetry broken JJ to be a fractional Josephson $\varphi_0$-junction \cite{SC:Buzdin2003}, which can exhibit a Josephson phase of $\varphi_0$ (neither 0 nor $\pi$) and a finite Josephson current in its ground state in the absence of any applied magnetic flux. The observation of this SOC-induced Josephson current would serve as a clear signal for topological superconductivity and MZMs. In addition, we show that the direction of the fractional Josephson current induced by the spin-phase is locked to the fermion parity of the topological JJ. It is noted that the SOC, driving non-trivial spin-phase-current relation, is inside the normal nanowire (black wire in Fig.~\ref{H-JJ}(a)), which is not coupled to a superconductor,  and thus readily tuned by an applied gate with the well developed spintronic technique \cite{Weber:2007_a,Koralek:2009_a,Wunderlich2010,Yang2012,Walser2012,Dettwiler2014}. We thus combine Majorana physics and spintronics, and propose an all-electrically controlled superconductor-semiconductor hybrid circuit to manipulate and control MZMs, and detect their non-Abelian braiding statistics \cite{TSC:Ivanov2001,TSC:Alicea2011,TSC:Beenakker2013a,TSC:Liu2014,TSC:Deng2015}.

Our work also shows the conceptual depth and complexity of semiconductor nanowire-based TSC \cite{TSC:Sau2010,TSC:Lutchyn2010,TSC:Oreg2010,TSC:Sau2010a} being well beyond the scope of the prototypical spinless $p$-wave TSC models (e.g. the Kitaev 1D model \cite{TSC:Kitaev2001}) since the spin degree of freedom plays no role in the latter type of manifestly spinless TSCs. The spintronic physics and the various spin-phase JJ physics being described in the current work simply do not exist within the 1D Kitaev (or for that matter, in any spinless $p$-wave TSC) model, showing conclusively that the topological superconductivity in the semiconductor-superconductor hybrid systems predicted in Refs.~\cite{TSC:Sau2010,TSC:Lutchyn2010} is much richer than and goes far beyond the simple spinless $p$-wave TSC model often used in the literature. The Majorana spintronics physics described in the current work arises entirely from the interplay among spin-orbit coupling, Zeeman splitting, and $s$-wave superconductivity in the semiconductor nanowire model, which leads not only to MZMs and TSC, but also to very rich  spin-phase JJ physics with manifest experimental consequences not present at all in the Kitaev model of spinless $p$-wave superconductivity.  It is interesting that all of the rich spin-phase JJ physics being discussed in the current work in the context of TSC systems shows up only in the realistic semiconductor nanowire systems and not at all in the idealized models of spinless $p$-wave TSC, perhaps explaining why this important subject has so far been mostly ignored in the literature.

We clarify here the notations for the various JJs to be used throughout the article. We use the  nomenclatures 0-junction (i.e. conventional JJ), $\pi$-junction, and $\varphi_0$-junction throughout this paper to refer to the JJs which exhibit the Josephson phase of $0$ \cite{SC:Josephson1962}, $\pi$ \cite{SC:Sigrist1992} and $\varphi_0$ (neither 0 or $\pi$) \cite{SC:Buzdin2003} in their ground states respectively in the absence of any applied magnetic flux. We use N-junction and U-junction to refer to the N-shape and U-shape geometries of the JJs shown in Fig.~\ref{D-current-phase}(a) and \ref{D-current-phase}(b) respectively.

Our paper is organized as follows. In Sec II, we demonstrate an intrinsic $\pi$-phase difference between the spin-triplet superconducting pairings enforced by MZMs localized at 
the opposite ends of a 1D TSC. This $\pi$-phase difference arises from neither a charge phase nor an orbital effect, and only exists in the topological superconducting regime at the boundaries between topological and trivial phases (i.e. at interfaces or wire ends). We further establish that this $\pi$-phase is indeed a spin-phase and can lead to the implementation of topological Josephson $0$- and $\pi$-junction with N-shape and U-shape geometries (Fig.~\ref{D-current-phase}(a) and (b)). In Sec III, we show that the Majorana coupling and spin-phase difference across the topological JJ can be tuned by a gate-voltage controllable SOC in the normal nanowire, connecting the two topological superconductors (a U-shape topological JJ as shown in Fig.~\ref{H-JJ}(a)), with exponential accuracy which leads to flux-free-control of the MZM coupling energy and Josephson current, and leads to a fractional Josephson $\phi_0$-junction. In particular, the direction of the spin-phase driven topological Josephson current can measure the Fermion parity of a topological JJ. In Sec IV, we propose an all-electrically controllable superconductor-semiconductor hybrid circuit to detect the non-Abelian nature of MZMs and present a discussion of the experimental feasibility of the propoased device. In Sec V, we conclude with a summary of our results and discussing possible future directions. (Some of the more complicated technical details are relegated to Appendices A and B although the results and equations from these appendices are sometimes used in the main text.)

\section{Spin state dependent $0-\pi$ Josephson junction transition}\label{0-pi-sec}

In a SC/NM junction, Cooper pairs can tunnel into the NM which inherits certain superconducting properties such as supercurrent. If the tunneling 
between the superconductor and NM obeys spin rotation SU(2) symmetry, 
the phase of the induced superconducting condensate in the NM is normally expected to only
depend on the superconducting charge phase and orbital-phase. 
However we find that the spin-phase, associated with the spin degree of freedom of Cooper pairs, may also play an important role in certain JJs, to be elaborated below.

We firstly consider the recently extensively studied time-reversal symmetry broken 1D TSC tight-binding model (a semiconductor nanowire with SOC coupling in the presence of a bulk superconductor and Zeeman spin splitting) whose Hamiltonian in the basis $(c_{\uparrow},c_{\downarrow},-c^{\dagger}_{\downarrow},c^{\dagger}_{\uparrow})^{\text T}$ has the form \cite{TSC:Sau2010,TSC:Lutchyn2010,TSC:Oreg2010,TSC:Sau2010a}
\begin{eqnarray}\label{Ham-SOC-1}
H_{\rm TS}&=&(-2t_s\cos(kd)-\mu_{\rm s})\tau_z\otimes \sigma_0-M \tau_0\otimes \sigma_z\nonumber \\
&+&2t_{\rm so} \sin(kd) \tau_z\otimes \sigma_y+\Delta \tau_x\otimes \sigma_0,
\end{eqnarray}
where $k$ and $d$ are the wave vector and the lattice constant respectively, $t_s$ is the spin independent hopping, $t_{\rm so}$ is the SOC strength, and $\mu_{\rm s}$ is the chemical potential of the semiconductor nanowire with $M$ the Zeeman coupling strength and $\Delta$ the proximity induced superconducting gap. In the topological superconducting regime of the semiconductor nanowire \cite{TSC:Sau2010,TSC:Lutchyn2010,TSC:Oreg2010,TSC:Sau2010a}, there are two MZMs $\gamma_{1}$ and $\gamma_{2}$ located at the right and left ends respectively (Fig.~\ref{D-current-phase}(a,b)).
In the strong Zeeman splitting limit, $M \gg \Delta$, the spin polarization of the two MZMs is almost anti-parallel to the magnetization. Besides, the Hamiltonian in Eq.~\eqref{Ham-SOC-1} commutes with the complex conjugation operator $\mathcal{K}$ so that MZMs should be eigenfunctions of $\mathcal{K}$. Thus, the two MZMs at right and left ends have the form
\begin{eqnarray*}
\gamma_{1\uparrow}=(c_{\uparrow}+c_{\uparrow}^{\dagger}),\ \ \gamma_{2\uparrow}=i(c_{\uparrow}-c_{\uparrow}^{\dagger}),
\end{eqnarray*}
which are even and odd under complex conjugation respectively.
Correspondingly, spin-triplet pairing coefficients \cite{SC:Leggett1975} are $\psi_{\uparrow\uparrow,1}=-\psi_{\uparrow\uparrow,2}=1$ (see details in appendix \ref{spin-triplet-pair}). The minus sign for $\psi_{\uparrow\uparrow,2}$ is from the square of the $i$ in $\gamma_2$. 
According to Eq.~\eqref{spin1-3} in our appendix \ref{spin-triplet-pair}, the anomalous density matrices \cite{TSC:Liu2015} for the MZMs at the two ends are  
\begin{eqnarray}\label{density-1}
f_{1}=\left(\begin{array}{cc} 0 & 1 \\ 0 & 0 \end{array}\right), \ \  f_{2}=\left(\begin{array}{cc} 0 & -1 \\ 0 & 0 \end{array}\right).
\end{eqnarray}  If the TSC respects time-reversal symmetry, we will have MZMs in the spin-down channel as well,  
\begin{eqnarray}\label{MZM-2}
\gamma_{1\downarrow}&=&\hat{T} \gamma_{1\uparrow}\hat{T}^{-1} =c_{\downarrow}+c_{\downarrow}^{\dagger},\nonumber \\ \gamma_{2\downarrow}&=&\hat{T} \gamma_{2\uparrow}\hat{T}^{-1}=i(c_{\downarrow}-c_{\downarrow}^{\dagger}),
\end{eqnarray}
with the time-reversal operator $\hat{T}=-i\sigma_y \mathcal{K}$.
The anomalous density matrices in this case are
\begin{eqnarray}\label{density-2} f_{\rm 1}= \left(\begin{array}{cc} 0 & 1 \\ 1 & 0 \end{array}\right) , \ \ f_{\rm 2}= -\left(\begin{array}{cc} 0 & 1 \\ 1 & 0 \end{array}\right) \end{eqnarray} with  $\psi_{\downarrow\downarrow,1}=-\psi_{\downarrow\downarrow,2}=1$. Based on Eqs.~(\ref{density-1}) and (\ref{density-2}), the MZM-induced spin-triplet superconducting condensates $f_{1,2}$ have a $\pi$-phase difference, 
regardless of whether time-reversal symmetry is broken or not. As MZM-induced pairing is odd-frequency $s$-wave spin-triplet \cite{TSC:Asano2013,TSC:Liu2015}, the $\pi$-phase difference arises from neither a charge phase nor an orbital-phase.
We emphasize here that this $\pi$-phase difference is not just a mathematical construct, but has observable physical effects as shown below.

In the time-reversal symmetry broken TSC/NM/TSC JJs (Fig.~\ref{D-current-phase}(a) and \ref{D-current-phase}(b)), the two identical nanowires (red wires) are proximity-induced topological superconductors. The normal-metal wire (black wire) connect the two TSCs in a different way which forms N- and U-junction as shown in Fig.~\ref{D-current-phase}(a) and \ref{D-current-phase}(b) respectively. The two TSC nanowires (red wires) are described in the minimal model by Eq.~(\ref{Ham-SOC-1}) and the normal wire (black wire) is described by
\begin{eqnarray}\label{Ham-SOC-2}
H_{\rm N}&=&(-2t_N\cos(kd)-\mu_{\rm N})\tau_z\otimes \sigma_0,
\end{eqnarray} where $t_{\rm N}$ and $\mu_{\rm N}$ are
the hopping energy and chemical potential in the normal-metal wire respectively. The MZMs $\gamma_{1}^{a}$ and $\gamma_{2}^{a}$ ($\gamma_{1}^{b}$ and $\gamma_{2}^{b}$) are 
located
at the left and right ends of the $a$ ($b$) wires (Fig.~\ref{D-current-phase}(a,b)). In the strong Zeeman splitting limit $M \gg \Delta$, the spin direction of these MZMs is antiparallel to the magnetic field \cite{TSC:He2014}.  
According to Eq.~(\ref{density-1}), the superconducting condensates induced by these two MZMs 
have
a $\pi$-phase difference. It follows that the current-phase relation of N-junction (Fig.~\ref{D-current-phase}(a)) has a $\pi$-phase shift 
as compared to the coupling  of $\gamma_{2}^{\rm a}$ and $\gamma_{2}^{\rm b}$ in the U-junction (Fig.~\ref{D-current-phase}(b)).
To test this prediction, we set the Hamiltonian parameters in Eq.~(\ref{Ham-SOC-1}) to be in the topological superconducting regime ($M > \Delta$, $\mu_s=0$) and plot the eigenenergies as a function of the charge phase difference $\phi$ in Figs~\ref{D-current-phase}(c) and \ref{D-current-phase}(d), which correspond to the N- and U-junction respectively. 
In Figs.~\ref{D-current-phase}(c) and \ref{D-current-phase}(d), the Andreev levels cross at $\phi=\pi$ and $\phi=0$, which indicates the 4$\pi$ periodic $0$ and $\pi$ JJs respectively. The lines with constant $E=0$ correspond to the MZMs at other uncoupled ends (Figs.~\ref{D-current-phase}(a) and \ref{D-current-phase}(b)).  
Moreover, we tune the superconductors into the topologically trivial regime by choosing $M=\Delta/2$ without changing other parameters and plot the eigenenergies of the N- and U-junction as a function of $\phi$ in Figs.~\ref{D-current-phase}(e) and \ref{D-current-phase}(f) respectively. The Andreev levels in both of the two JJs behave like the normal Josephson $0$-junction whose minimal ground state is at $\phi=0$ with 2$\pi$ periodicity. Besides, we also calculate the eigenenergies for $M=2\Delta,4\Delta, 6\Delta$ (topologically nontrivial regime) and $M=0,0.3\Delta,0.6\Delta$ (topologically trivial regime) in the two JJs. 
We find that in the topological trivial regime, the N-shape and U-shape junctions always have the same Andreev levels with a normal Josephson $0$-junction with $2\pi$ periodicity. However, in the topologicall nontrivial regime, besides the arising of the $4\pi$ periodicity of both N-shape and U-shape, the U-junction becomes $\pi$-junction with MZMs located at the interface of the TSC/NM interface. These results confirm that the appearance of $\pi$-phase in the U-junction only depends on the presence of the MZM-induced spin-triplet pairings. Thus this $\pi$-phase is indeed a spin-phase. 

We also consider the time-reversal invariant TSC/NM/TSC junction whose TSC Hamiltonian is given as \cite{TSC:Zhang2013}
\begin{eqnarray}\label{gap-iron}
H_{\rm TS}&=&(-2t_s\cos(kd)-\mu_{\rm s})\tau_z\otimes \sigma_0 \nonumber \\
&&+2t_{\rm so} \sin(kd) \tau_z\otimes \sigma_z+\Delta(kd) \tau_x\otimes \sigma_0,
\end{eqnarray}
where $\Delta(kd)=(\Delta_0-\Delta_1\cos(kd))$. 
This spin-singlet superconducting gap involves both $s_+$ and $s_-$ channels, and it vanishes at $\cos(k_0d)=\Delta_0/\Delta_1$.
The SOC in the semiconductor nanowire induces the spin splitting, and leads to two Fermi wave vectors $k_{\rm 1f}$ and $k_{\rm 2f}$ with $k_{\rm 1f}<k_{\rm 2f}$. For $k_{\rm 1f}<k_0<k_{\rm 2f}$,
the system is in a topological superconducting regime~\cite{TSC:Zhang2013}, and the associated Andreev levels of the N- and U-junction are similar to those plotted in Figs.~\ref{D-current-phase}(c) and \ref{D-current-phase}(d) 
except that there is a 
Kramers degeneracy in this time-reversal symmetric case. Taking $\Delta_1=0$, the system is in the topologically trivial regime and the Andreev levels for the two JJ configurations are similar to those plotted in Fig.~\ref{D-current-phase}(e) and \ref{D-current-phase}(f) 
except for the Kramers degeneracy. 
We thus conclude that for the time-reversal symmetric topological JJ, the superconducting condensates at the opposite ends have a $\pi$-phase difference provided that there exist MZM-induced spin-triplet pairings. The $\pi$-phase is related to the Cooper pair spin-triplet states, and thus belongs to the spin-phase which is similar to the time-reversal symmetry broken case.

In a real semiconductor nanowire, which has a finite width in its transverse plane ($x$-$y$ plane), 
the complex conjugation symmetry is broken in general, for example, by considering a SOC form $-i\partial_y \sigma_z$. 
However, even in this case, we can still establish the MZM-related spin-phase difference in topological JJs, 
by theoretically treating the system as a multi-band TSC, in the presence of the following mirror reflection symmetry 
\begin{eqnarray*}
&&\mathcal{M}_{z} H(k_x,k_y,k_z,\sigma_x,\sigma_y,\sigma_z)\mathcal{M}_z^{-1} \nonumber \\
&&=H(k_x,k_y,-k_z,-\sigma_x,-\sigma_y,\sigma_z).
\end{eqnarray*}
In Appendix~\ref{0-pi}, we show that the $s$-wave spin-triplet Cooper pairs described by a $\vec{d}$-vector (which is a vector description of spin-triplet superconducting condensates introduced in Ref. \cite{SC:Leggett1975} as defined in Eq.~\eqref{d-vector-1} in our appendix \ref{spin-triplet-pair}) along the $x$ or $y$ direction 
are odd under the mirror reflection, $\mathcal{M}_z$. By contrast, the $s$-wave Cooper pairs of spin-singlet or spin-triplet with the $\vec{d}$-vector along the $z$ direction
are even. Thus, with a $\vec{d}$-vector along the $x$ or $y$ direction, 
the MZM-induced Cooper pairing at the two ends of the TSC are always opposite in sign, and thus have a $\pi$-phase difference, provided that the system respects the mirror reflection symmetry.
 
\section{SOC tunable Majorana fermion couplings and unconventional Josephson effects}
Inspired by the $\pi$ spin-phase and its induced topological Josephson $\pi$-junction, we expect to have a completely different technique to manipulate the Majorana fermion (MF) coupling in topological JJs from the method of using magnetic flux to control the phase of charge origin \cite{SC:Golubov2004a}. 
Because SOC, with the general form $(\bm{\nabla} V \times \bm{\sigma})\cdot \bm{p}$, performs 
as a spin-dependent vector potential and can rotate the $\vec{d}$-vector of spin-triplet Cooper pairs \cite{SC:Bergeret2013,SC:Liu2014,SC:Bergeret2014}, its effect on the spin-phase will affect the MF coupling and the current-phase relation in a topological JJ.

\begin{figure}[ht]
\centering
\begin{tabular}{l}
\includegraphics[width=0.8\columnwidth]{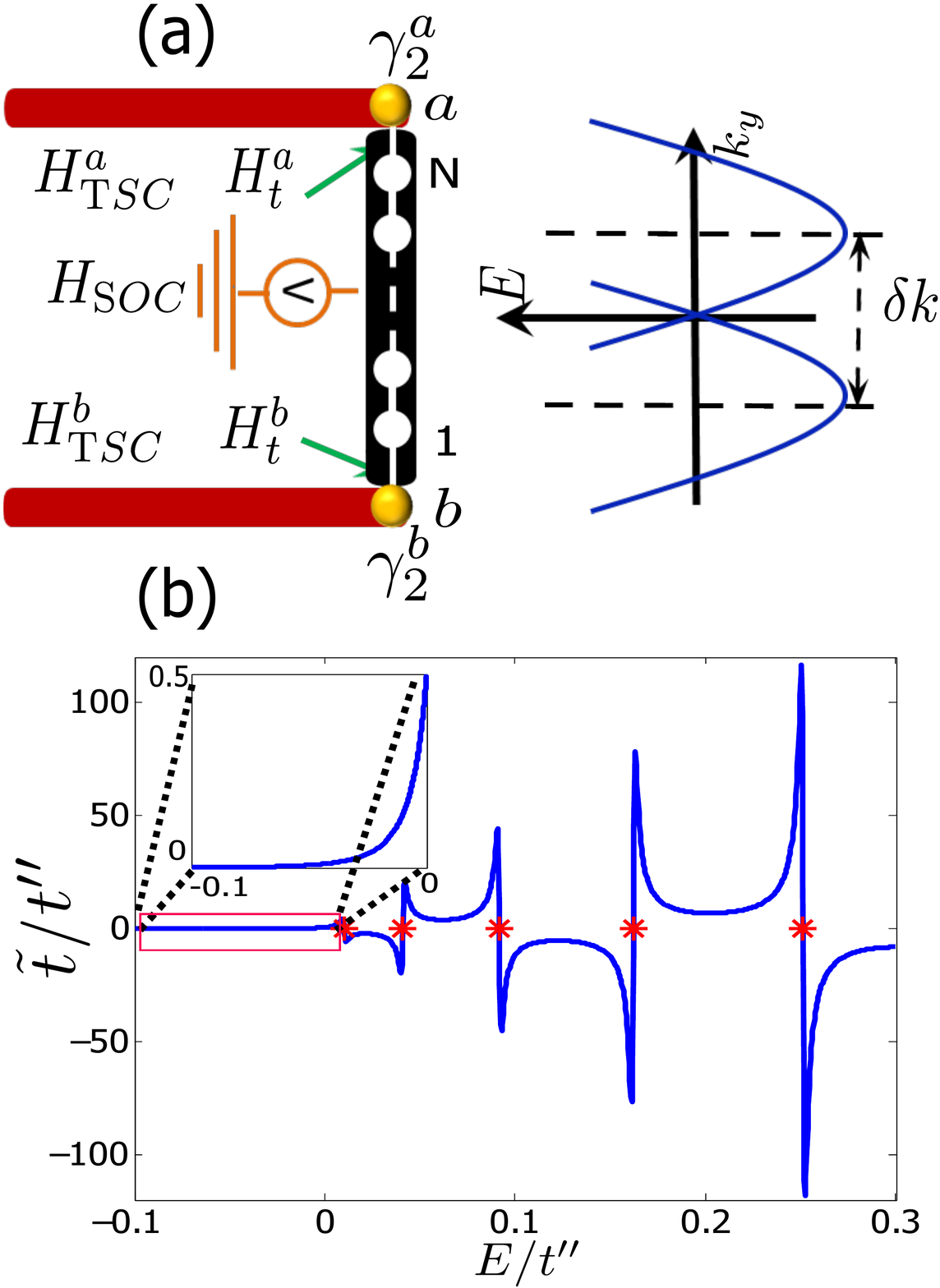}
\end{tabular}
\caption{(a) TSC/SOC-semiconductor/TSC junction. There are N sites (white dots) in the SOC wire (black wire). The red wires represent two TSCs. The yellow balls represent MZMs. The applied gate voltage can control the chemical potential and SOC inside the normal wire through standard technologies of the semiconductor spintronics. (b) The effective coupling amplitude $\tilde{t}$ as a function of energy $E$. Both horizontal and vertical axises are normalized by the hopping constant $t\rq{}\rq{}$ of the SOC wire. The red stars indicate the eigenenergy $\epsilon_n=2\sqrt{t''^2+t_{\rm so}^2}(1-\cos (k_nd))$ in the horizontal axis. The inset shows $\tilde{t}$ with the energy inside the semiconductor band gap.}
\label{H-JJ}
\end{figure}

To explore the SOC effect in a topological JJ, we consider the U-junction with the SOC normal wire along $y$ direction (Fig.~\ref{H-JJ}(a)). The Hamiltonian of this junction takes a general form
\begin{eqnarray}\label{Ham-TSC-SOC}
H=\left(\begin{array}{ccc} H_{\rm TSC}^{\rm a} & 0 & H_t^{\rm a} \\ 0 & H_{\rm TSC}^{\rm b} & H_t^{\rm b} \\ (H_t^{\rm a})^{\dagger} & (H_t^{\rm b})^{\dagger} & H_{\rm SOC} \end{array}\right).
\end{eqnarray} 
In this section, we consider both cases with and without time-reversal symmetry in the Hamiltonian $H_{\rm TSC}^{\rm a,b}$.

The Hamiltonian of the normal SOC wire $H_{\rm SOC}$ reads,
\begin{eqnarray*}
H_{\rm SOC}&=& \sum_{i,\sigma} \mu' c^{\dagger}_{i\sigma}c_{i\sigma} + \sum_{i,\sigma}-t'c^{\dagger}_{i+1\sigma}c_{i\sigma} \nonumber \\ &+& \sum_{i,\sigma}i(-1)^{\sigma}t'_{\rm so}c^{\dagger}_{i+1\sigma}c_{i\sigma}+\text{h.c.}, 
\end{eqnarray*}
with $\mu'$ the chemical potential, $t'$ the spin-independent hopping, and $t_{\rm so}'$ the strength of the experimentally accessible $p_y\sigma_z$ type SOC \cite{Weber:2007_a,Koralek:2009_a,Wunderlich2010,Yang2012,Walser2012,Dettwiler2014} which is tunable by a gate voltage as shown in Fig.~\ref{H-JJ}(a).

The couplings between the normal wire and superconductors in Eq.~(\ref{Ham-TSC-SOC}) are
\begin{eqnarray*}\label{Ham-cop-1}
H_{t}^{a}=\sum_{\sigma}t''c^{\dagger}_{a\sigma}c_{1\sigma}+\text{h.c}, \ \  H_{t}^{b}=\sum_{\sigma}t''c^{\dagger}_{b\sigma}c_{N\sigma}+\text{h.c}. 
\end{eqnarray*}
Here  the indices $1$ and $N$ indicate the sites at the opposite ends of
the SOC wire (Fig.~\ref{H-JJ}(a)), and the coupling $t''$ models 
the electron tunneling across the normal and superconducting wires.  
According to Eq.~(\ref{Ham-TSC-SOC}), the two TSCs are indirectly coupled through the SOC in the normal wire. To analyze the SOC effect in the TSC/SOC-semiconductor/TSC junctions, we derive an effective coupling between the two TSCs.  
The normal SOC wire contributes an effective self energy
\begin{eqnarray}\label{SE-1}
\Sigma=-H_{t}G_{\rm SOC}H^{\dagger}_{t},
\end{eqnarray}  with  $H_{t}=(H_{t}^{a},H_{t}^{b})^{\rm T}$ and the Green's function $G_{\rm SOC}$ given by  
\begin{eqnarray}\label{SE-2}
G_{\rm SOC}(y,y')=\sum_n\frac{|\psi_n(y)\rangle \langle \psi_n(y')|}{E-\epsilon_n+i\delta}. 
\end{eqnarray}
Here $\epsilon_n$ is the eigenenergy, and $|\psi_n\rangle$ is the eigenfunction of $H_{\rm SOC}$, which satisfies the boundary conditions
$\psi_n(0)=0$ and $\psi_n((N+1)d)=0$ with $d$ the lattice constant and $N$ the number of lattice sites in the normal SOC wire. 
Due to the SOC, the two spin bands will be shifted oppositely in $k$ axis (Fig.~\ref{H-JJ}(a))
by $\delta k=2\arcsin(t'_{\rm so}/\sqrt{t^2+t_{\rm so}^{'2}})/d$. 
The eigenfunctions for the two spin channels take the form 
\begin{eqnarray}\label{SE-3}
\psi_{k,\uparrow}(y)&=&\left(\begin{array}{c} 1 \\ 0 \end{array}\right)e^{i\frac{\delta k }{2}y}\sin(ky), \nonumber \\
\psi_{k,\downarrow}(y)&=&\left(\begin{array}{c} 0 \\ 1 \end{array}\right)e^{-i\frac{\delta k}{2}y}\sin(ky)
\end{eqnarray}
with $k_n=n\pi/(N+1)d$ according to the boundary condition.
By integrating out the electron and hole degrees of freedom in the normal nanowire, we obtain The off-diagonal term
\begin{eqnarray}\label{Ht-1}
&&\Sigma_{ab}=\Sigma_{ba}^{\dagger}=\tilde{t}e^{i\frac{\delta k(N-1)d}{2}\sigma_z}=\tilde{t}e^{i\frac{\theta}{2}\sigma_z},\nonumber \\
&&\tilde{t}=t''^2\sum_k\frac{\sin(ka)\sin(kNd)}{E-2\sqrt{t''^2+t_{\rm so}^2}(1-\cos(kd))},
\end{eqnarray} which gives rise to an effective spin-phase dependent coupling between two TSCs, $\tilde{t}e^{i\frac{\theta}{2}\sigma_z}$. 
In Eq.~\eqref{Ht-1}, we neglect the contribution from the poles of $G_{\text soc}$ in Eq.~(\ref{SE-3}) if there is no eigenstate inside the superconducting gap. This is valid when the chemical potential of the SOC wire is in its semiconductor band gap, or the length of the SOC wire $L=(N+1)d$ is much smaller than the coherence length $\xi$ so that around the Fermi surface, $\epsilon_{n+1}-\epsilon_n \gg \Delta$. We plot $\tilde{t}$ as a function of energy $E$ in Fig.~\ref{H-JJ}(b). The effective coupling amplitude changes sign with the energy $E$ across the quantized eigenenergy $\epsilon_n$ and becomes a pure exponential decay inside the semiconductor band gap. If we also consider the charge phase by adding a vector potential $(A,0,0)^{\text T}$ in the SOC region, the tunneling Hamiltonian takes the form
\begin{eqnarray}\label{Ht-2}
\tilde{H}_t=\tilde{t}(c^{\dagger}_a e^{\frac{i}{2}( \phi \sigma_0 + \theta \sigma_z)} c_{b}- c_{a} e^{-\frac{i}{2}( \phi \sigma_0 + \theta \sigma_z)}c^{\dagger}_b) 
\end{eqnarray}
where 
\begin{eqnarray}\label{phase-1}
\phi/2=eAL/\hbar, \ \ \theta=\delta k L,
\end{eqnarray} correspond to the charge phase and spin-phase respectively for an electron traveling across the junction. 

\begin{figure}[htbp]
\centering
\begin{tabular}{l}
\includegraphics[width=0.8\columnwidth]{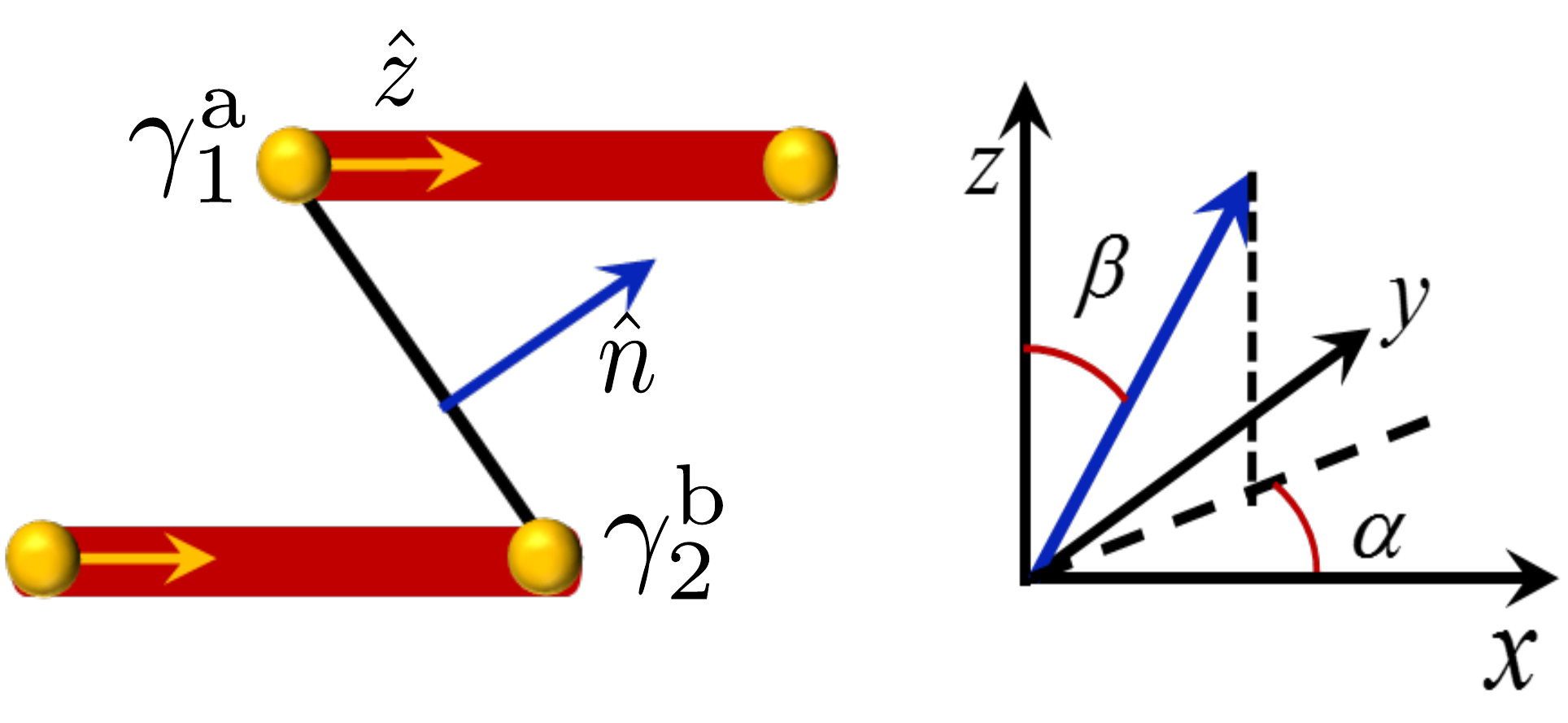}
\end{tabular}
\caption{A schematic plot of the TSC/SOC-semiconductor/TSC junction with misalignment between the spin quantization axis and SOC field direction. The yellow and blue arrows indicate the MF spin direction $\hat{\bm{z}}$ and SOC field direction $\hat{\bm{n}}$. Given $\hat{z}$ along $z$ axis, the SOC field direction can be described by the polar angle $\beta$ and azimuthal angle $\alpha$.}
\label{arbitrary-d}
\end{figure}

The effective coupling Hamiltonian, Eq.~(\ref{Ht-2}), can be easily generalized to the case for the arbitrary SOC field direction $\bm{\hat{n}}$ (Fig.~\ref{arbitrary-d}) as \begin{eqnarray}\label{Ht-2-1}
&&\tilde{H}_t=\tilde{t}(c^{\dagger}_a e^{\frac{i}{2}( \phi \sigma_0 + \theta \bm{\hat{n}}\cdot \bm{\sigma})} c_{b}- c_{a} e^{-\frac{i}{2}( \phi \sigma_0 + \theta \bm{\hat{n}}\cdot \bm{\sigma}^*)}c^{\dagger}_b) 
\end{eqnarray}
As we are interested in the topological JJ, it is convenient to write the tunneling Hamiltonian in the Majorana representation by a unitary transformation \cite{TSC:Beenakker2015}
\begin{widetext}
\begin{eqnarray}\label{Ht-20}
\tilde{H}_t&=& \frac{\tilde{t}}{4} \left(\begin{array}{c}\gamma^{a}_{1} \\ \gamma^{a}_{2} \end{array}\right)^{\text T} \left(\begin{array}{cc} e^{\frac{i}{2}( \phi \sigma_0 + \theta \bm{\hat{n}}\cdot \bm{\sigma})}-e^{-\frac{i}{2}( \phi \sigma_0 + \theta \bm{\hat{n}}\cdot \bm{\sigma}^*)} & i \left( e^{\frac{i}{2}(\phi \sigma_0 + \theta \bm{\hat{n}}\cdot \bm{\sigma})}+ e^{-\frac{i}{2}( \phi \sigma_0 + \theta \bm{\hat{n}}\cdot \bm{\sigma}^*)}\right) \\ -i \left(e^{\frac{i}{2}( \phi \sigma_0 + \theta \bm{\hat{n}}\cdot \bm{\sigma})}+ e^{-\frac{i}{2}( \phi \sigma_0 + \theta \bm{\hat{n}}\cdot \bm{\sigma}^*)}\right) & e^{\frac{i}{2}(\phi \sigma_0 + \theta \bm{\hat{n}}\cdot \bm{\sigma})}-e^{-\frac{i}{2}( \phi \sigma_0 + \theta \bm{\hat{n}}\cdot \bm{\sigma}^*)} \end{array}\right) \left(\begin{array}{c}\gamma^{b}_{1} \\ \gamma^{b}_{2} \end{array}\right), \nonumber \\
\nonumber \\
 &&\left(\begin{array}{c}\gamma_{1} \\ \gamma_{2} \end{array}\right)=\sqrt{2}U \left(\begin{array}{c}c \\ c^{\dagger} \end{array}\right),\ \ U= \frac{1}{\sqrt{2}} \left(\begin{array}{cc} \sigma_0 & \sigma_0 \\ -i\sigma_0 & i\sigma_0 \end{array}\right), 
\end{eqnarray}
with $\gamma_{1}^a=(\gamma_{1\uparrow},\gamma_{1\downarrow})^{\text T}$, $\bm{\hat{n}}\cdot\bm{\sigma}=\cos\beta\sigma_z+\sin\beta\cos\alpha\sigma_x +\sin\beta\sin\alpha\sigma_y$, $\beta$ the polar angle and $\alpha$ the azimuthal angle (Fig~\ref{H-JJ}(b)). Equation~(\ref{Ht-20}) provides 
a general form of coupling
between two TSCs through a SOC wire. In the following, we use it to study unconventional Josephson effects in both time-reversal broken and invariant junctions. We refer to this description as the MF representation.
\end{widetext}

\subsection{Time-reversal broken topological Josephson junction}\label{TRB-TJJ}
For a time-reversal symmetry broken TSC/SOC-semiconductor/TSC junction, we assume (without loss of generality) that the MZMs at sites $a$ and $b$ are $\gamma^{a}_{1\uparrow}$ and $\gamma^{b}_{2\uparrow}$ respectively (Fig.~\ref{H-JJ}(a)). Then according to Eq.~(\ref{Ht-20}), the MF coupling Hamiltonian reads
\begin{eqnarray}\label{Ht-8}
\tilde{H}_t=\frac{i\tilde{t}}{2}(\cos\frac{\theta}{2}\cos\frac{\phi}{2}-\sin\frac{\theta}{2}\sin\frac{\phi}{2}\cos\beta)\gamma_{1\uparrow}^{a}\gamma_{2\uparrow}^{b}.
\end{eqnarray}
The associated Andreev levels and Josephson currents are
\begin{subequations}
\begin{align}
E=\pm\frac{\tilde{t}}{2}(\cos\frac{\theta}{2}\cos\frac{\phi}{2}-\sin\frac{\theta}{2}\sin\frac{\phi}{2}\cos\beta) \label{AL-5},\\
I=\mp\frac{\tilde{t}e}{\hbar}(\cos\frac{\theta}{2}\sin\frac{\phi}{2}+\sin\frac{\theta}{2}\cos\beta\cos\frac{\phi}{2})\label{JC-5}.
\end{align}
\end{subequations}
For $\beta=\frac{\pi}{2}$, the MF spins are perpendicular to the $z$ axis and the Andreev levels take the form
\begin{equation}
E=\pm\frac{\tilde{t}}{2}\cos\frac{\theta}{2}\cos\frac{\phi}{2}\label{AL-5-1}. 
\end{equation}
This indicates that the Andreev level crossing is always at $\phi=\pi$, the minimal ground state energy is always at $\phi=0$, and the topological JJ is always a Josephson 0-junction.
However the MF coupling energy 
oscillates as a function of the spin-phase $\theta$, which is consistent with our previous study \cite{TSC:Liu2015}. 

\begin{figure}[htbp]
\centering
\begin{tabular}{l}
\includegraphics[width=0.8\columnwidth]{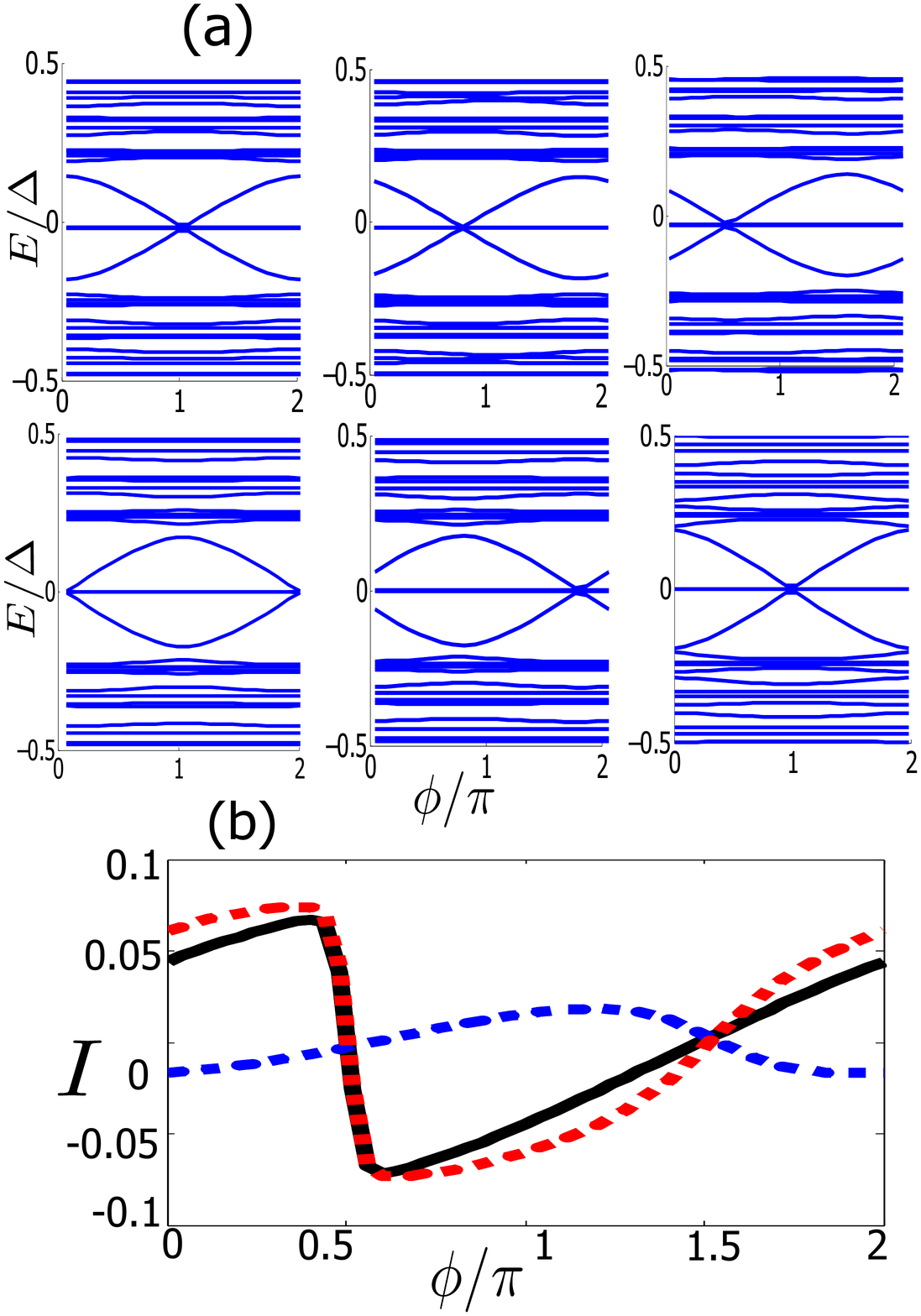}
\end{tabular}
\caption{(a) Andreev bound states with effective magnetic field of SOC perpendicular to the d-vector of the MF at the end of TSCs. The horizontal axis is the phase difference of the two TSCs and the vertical axis is energy. The panels corresponds to $\theta=0,\frac{\pi}{4}, \frac{\pi}{2}, \pi, \frac{5\pi}{4}$ and $2\pi$ respectively. (b) Josephson currents with the spin-phase $\theta=\pi/2$. The red dashed line represents the contribution from the MF coupling. The blue dashed line represent the contribution from the bulk superconducting ground state below the gap. The black solid line is the net Josephson current.}
\label{phi-Andreev}
\end{figure}

When the SOC field direction $\bm{\hat{n}}$ in the normal region is parallel to $\bm{\hat{z}}$ (along the two TSC wires (Fig.~\ref{arbitrary-d}))
so that $\beta=0$, the MF coupling Hamiltonian is
\begin{eqnarray}\label{Ht-4}
\tilde{H}_{\gamma}&=& \frac{\tilde{t}}{2} i\gamma_{1\uparrow}^{a}\gamma_{2\uparrow}^{b}\cos(\frac{\phi+\theta}{2}).
\end{eqnarray}
The associated Andreev levels and Josephson currents have the form
\begin{eqnarray}\label{AL-10}
E&=&\pm\frac{\tilde{t}}{2}\cos(\frac{\phi+\theta}{2}), \nonumber \\
I&=&\pm\frac{|e|\tilde{t}}{h}\sin(\frac{\phi+\theta}{2}),
\end{eqnarray}
which exhibit $4\pi$ periodicity in both charge phase $\phi$ and spin-phase $\theta$.
Here $+(-)$ indicates the fractional Josephson current direction and is related to the fermion parity defined as $ i\gamma_{1\uparrow}^{a}\gamma_{2\uparrow}^{b}=1-2c^{\dagger}c$ with $c=\gamma_{1}^a-i\gamma_2^b$.
The minimal ground state energy is shifted to $\phi=-\theta$. The obtained analytic results for the fractional Josephson relation  have  been confirmed in our numerical calculations (see Fig.~\ref{phi-Andreev}). 
It is worth emphasizing that even given charge phase $\phi=0$, we could have a finite spin-phase driven fractional Josephson current whose direction measures the fermion parity in the topological JJ.   

\begin{figure}[ht]
\centering
\begin{tabular}{l}
\includegraphics[width=0.9\columnwidth]{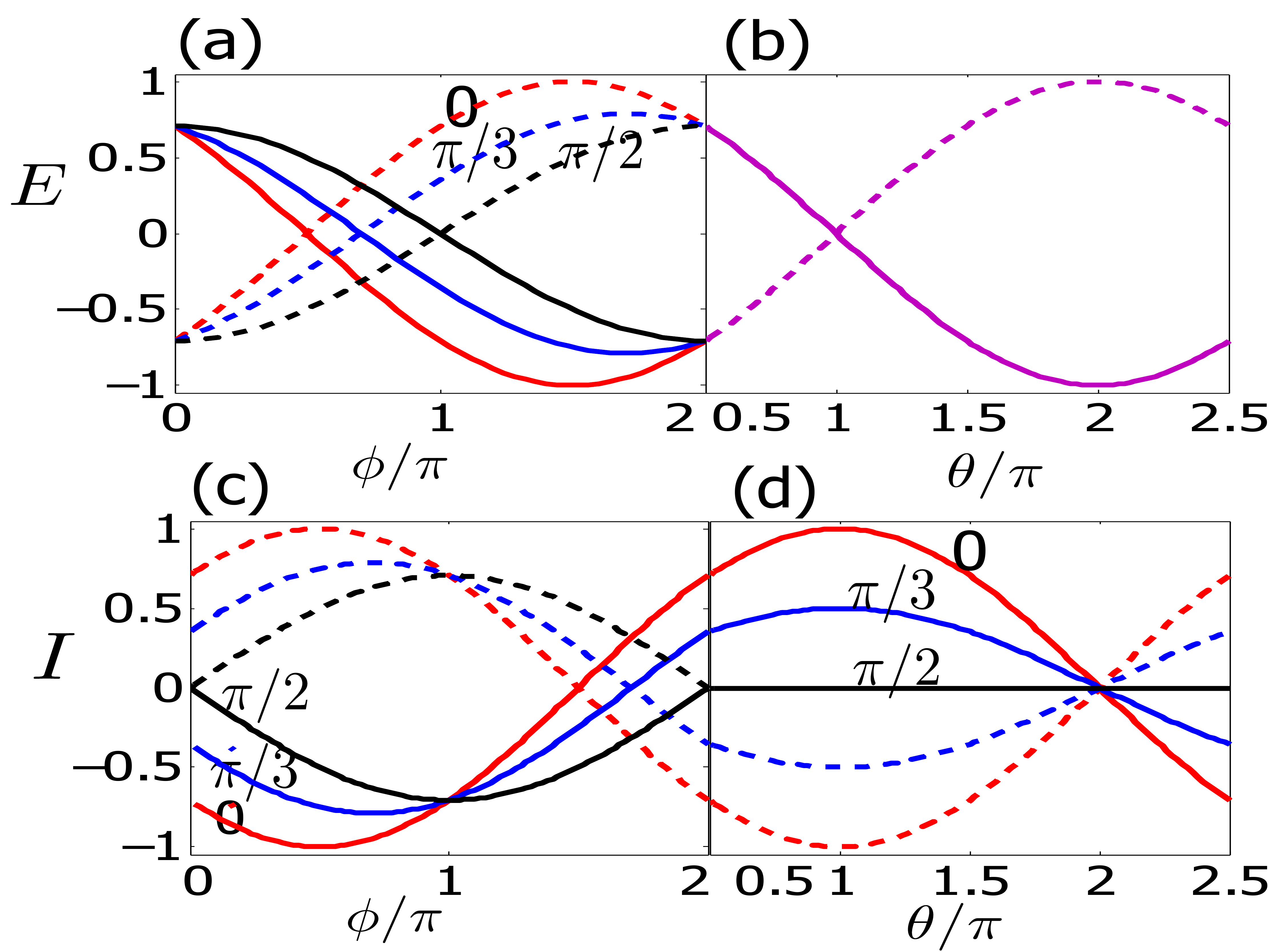}
\end{tabular}
\caption{ (a) and (c) The plot of Andreev levels and Josephson currents as a function of charge phase $\phi$ with given spin-phase $\theta=\pi/2$ and $\phi \in [-2\pi,0]$. The red, blue and black colors correspond to the misaligned angle $\beta=0,\pi/3,\pi/2$ respectively.  Energy and current are normalized by $\tilde{t}/2$ and $e\tilde{t}/h$ respectively with $e$ the electron charge and $h$ the Plank constant. The solid (dashed) curves represent the $+(-)$ in Eq.~(\ref{I-1}). (b) and (d) The plot of Andreev levels and Josephson currents as a function of spin-phase $\theta$ with fixed $\phi=0$ and $\theta \in [\pi/2,5\pi/2]$. In (b), the Andreev levels for $\beta=0,\pi/3,\pi/2$ collapse into the identical curves. }
\label{D-misalinged}
\end{figure}

For an arbitrary SOC field direction, we find that the minimal ground state energy is generally shifted away from $\phi=0$ due to the spin-phase $\theta$ unless $\beta=0$ as shown in Fig.~\ref{D-misalinged}(a). 
Consequently, the corresponding Josephson current has both $\cos(\phi/2)$ and $\sin(\phi/2)$ terms according to Eq.~(\ref{JC-5}) and therefore can be finite even at $\phi=0$ (Fig.~\ref{D-misalinged}(c)). In Figs.~\ref{D-misalinged}(a) and \ref{D-misalinged}(c),  the spin-phase is set to be $\pi/2$, 
for which the SOC effect leads to two terms of equal weight in the right hand sides of Eqs.~(\ref{AL-5},\ref{JC-5}).

\begin{figure}[ht]
\centering
\begin{tabular}{l}
\includegraphics[width=0.8\columnwidth]{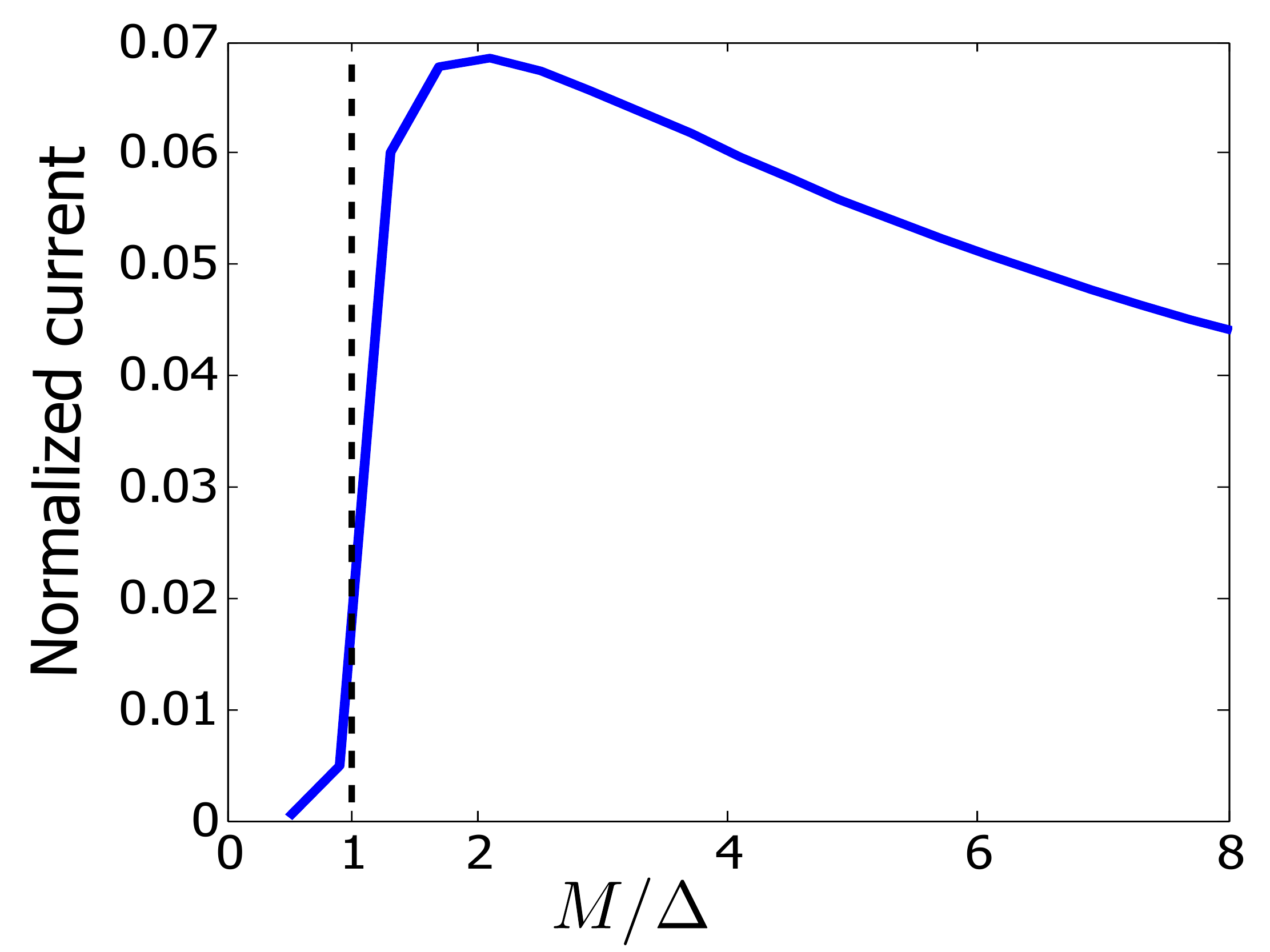}
\end{tabular}
\caption{Josephson current as a function of the magnetization with spin-phase $\theta=\pi/2$ and charge phase $\phi=0$. $M/\Delta=1$ is the topological phase transition point.}
\label{cvm}
\end{figure}

To explicitly illustrate this unusual Josephson current phenomena, we focus on $\phi=0$ and study the spin-phase-current relation. In this case,  
Josephson current has the form
\begin{eqnarray*}\label{I-1}
I=\mp\frac{\tilde{t}e}{\hbar}\sin\frac{\theta}{2}\cos\beta,
\end{eqnarray*}
which remains the $4\pi$ periodicity for arbitrary SOC field direction.
We plot the above spin-phase dependent Andreev levels and Josephson currents in Fig~\ref{D-misalinged}(b,d) for $\beta=0,\pi/3,\pi/2$. The solid (dashed) lines correspond to $+(-)$ sign in Eq.~(\ref{I-1}). At $\phi=0$, unless the polar angle $\beta=\pi/2$, the Josephson current can be turned on and off by varying the spin-phase (Fig.~\ref{D-misalinged}(d)), 
and we have a $\varphi_0$-junction as introduced in Ref. \cite{SC:Buzdin2003} in a completely different context. To understand the appearance of this $\varphi_0$-junction in a time-reversal symmetry  broken topological JJ, we compare the charge phase with the spin-phase in the coupling Hamiltonian Eq.~(\ref{Ht-2-1}). For $\hat{\bm{n}}\parallel \hat{\bm{z}}$, the Hamiltonian in Eq.~(\ref{Ht-2-1}) is diagonal in spin space. If we focus on the spin-up channel, the spin-phase plays exactly the same role as the charge phase so that it can be used to control the Josephson current. For $\hat{\bm{n}}\perp \hat{\bm{z}}$, spin-phase performs like the Zeeman coupling which rotates the MF spin direction without introducing a relative phase between the sites ``$a$" and ``$b$" (Fig.~\ref{H-JJ}(a)).
In this case, Josephson current cannot be turned on solely by the spin-phase. Thus, as long as the SOC field has a finite component along $\hat{\bm{z}}$ direction, this component will turn on the supercurrent in the topological JJ even at $\phi=0$. When the superconductor is in the topologically trivial regime, say $0\le M <\Delta$ for $\mu=0$, the magnetization induced bulk spin-triplet pairing has no spin polarization \cite{SC:Linder2015} and thereby the spin-triplet Cooper pairs have the same amplitude in both spin-up and spin-down channels. Consequently, the Josephson current in the topologically trivial regime should be zero at $\phi=0$. In Fig.~\ref{cvm}, we plot the Josephson current as a function of magnetization $M$ with $\theta=\pi/2$ and $\phi=0$ for the topological Josephson junction illustrated in Fig.~\ref{H-JJ}(a). The Josephson current drops down to zero sharply at the topological quantum phase transition point. The observation of this SOC-induced Josephson current would serve as a clear signal for topological superconductivity and MZMs. On the other hand, as the direction of the SOC-induced fractional Josephson current is solely determined by the fermion parity of the topological Josephson junction if the sign of the coupling amplitude $\tilde{t}$ is fixed, this property can be used to detect the non-Abelian nature of MZMs which will be discussed in the next section (Sec. \ref{non-Abelian}). 

\subsection{Time-reversal invariant topological Josephson junctions}

\begin{figure}[ht]
\centering
\begin{tabular}{l}
\includegraphics[width=0.8\columnwidth]{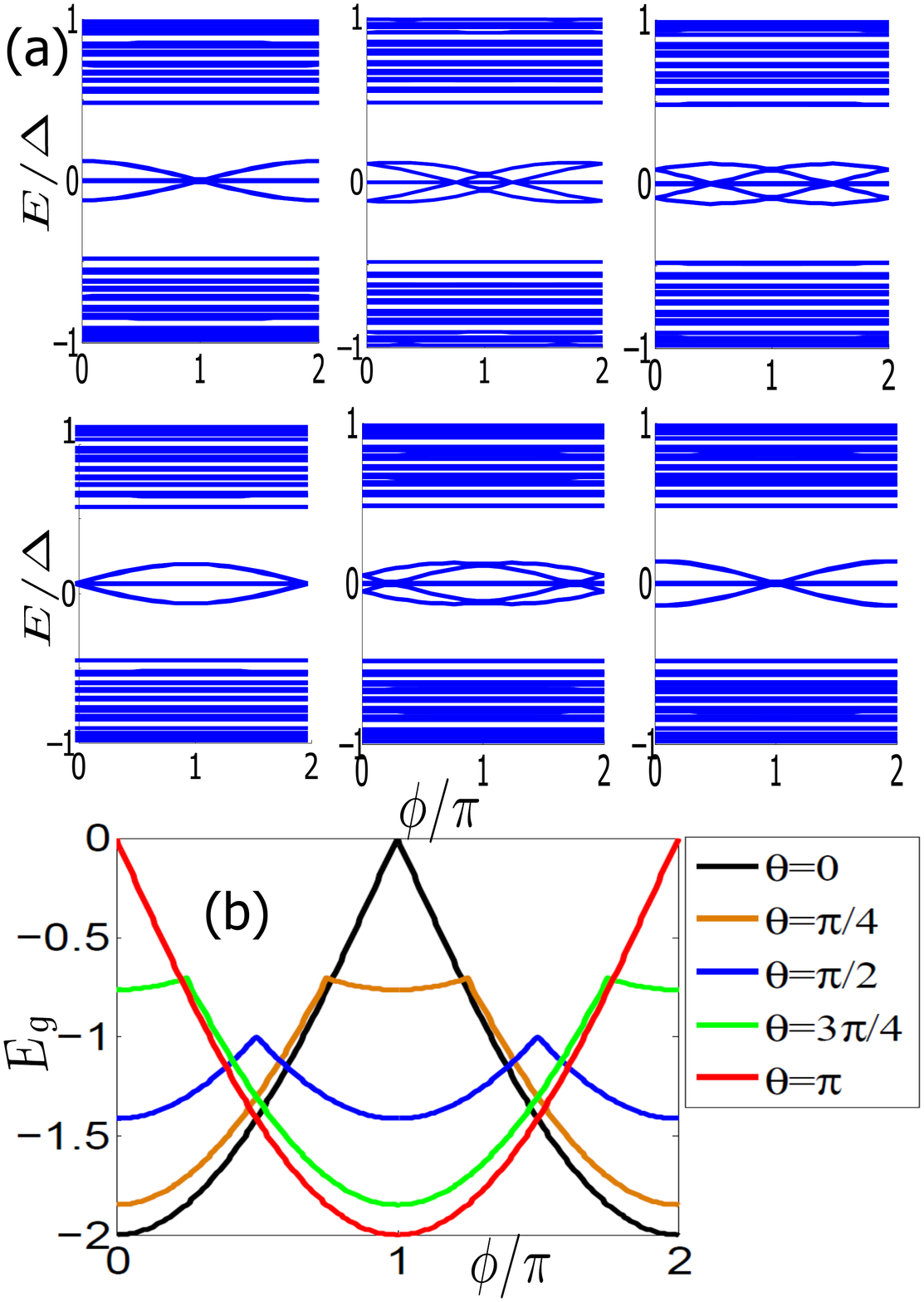}
\end{tabular}
\caption{ (a)Andreev bound states when the effective magnetic field of SOC is parallel to the MF spins at the end of TSCs. The horizontal axis is the phase difference of the two TSCs and the vertical axis is energy. The panels corresponds to $\theta=0,\frac{\pi}{4}, \frac{\pi}{2}, \pi, \frac{5\pi}{4}$ and $2\pi$ respectively. (b) The ground state energy $E_g$ as a function of charge phase $\phi$ given spin-phase $\theta=0,\pi/4,\pi/2,3\pi/4,\pi$. The ground state energy $E_g$ is normalized by $\tilde{t}/2$. }
\label{misalinged}
\end{figure}

In the low energy limit, $E \ll \Delta$, we consider the time-reversal invariant TSC/SOC-semiconductor/TSC JJ with a pair of MZMs $(\gamma^{a(b)}_{1(2)\uparrow},\gamma^{a(b)}_{1(2)\downarrow})^{\text T}$ located at site $a(b)$ (Fig.~\ref{arbitrary-d}). For simplicity, we first assume that the SOC field direction is parallel to $\hat{\bm{z}}$ axis. According to Eq.~(\ref{Ht-20}), the tunneling Hamiltonian can be projected to the Hilbert space expanded by these four MZMs as
\begin{eqnarray}\label{Ht-3}
\tilde{H}_{\gamma}&=& \frac{\tilde{t}}{4}\gamma_{1}^a  i \left( e^{\frac{i}{2}(\phi \sigma_0 + \theta \sigma_z)}+ e^{-\frac{i}{2}( \phi \sigma_0 + \theta \sigma_z)}\right) \gamma_{2}^b \nonumber \\
 &=&\frac{\tilde{t}}{2} i\gamma_{1\uparrow}^{a}\gamma_{2\uparrow}^{b}\cos(\frac{\phi+\theta}{2})+\frac{\tilde{t}}{2}i\gamma_{1\downarrow}^{a}\gamma_{2\downarrow}^{b} \cos(\frac{\phi-\theta}{2}).
\end{eqnarray}
According to Eq.~(\ref{Ht-3}), the MF coupling is not only determined by the charge phase $\phi$ but also depends on the spin-phase $\theta$. The associated Andreev levels and Josephson current have the form
\begin{subequations}
\begin{align}
E=\pm\frac{\tilde{t}}{2}\cos(\frac{\phi\pm\theta}{2}), \label{AL-3} \\
I=\mp\frac{2e\tilde{t}}{h}\sin\frac{\phi}{2}\cos\frac{\theta}{2}. \label{JC-3}
\end{align}
\end{subequations}
To confirm our analytical results, we numerically calculate the eigenenergies of the time-reversal invariant TSC/SOC-semiconductor/TSC junction (see Fig.~\ref{misalinged}(a)).
The spin-phase $\theta$ shifts the two branches of Andreev levels oppositely in the $\phi$ axis (Fig~\ref{misalinged}(a)), which is consistent with our analytical results in Eq.~(\ref{AL-3}). Based on Eq.~(\ref{AL-3}), the ground state energy takes the form 
\begin{eqnarray}\label{E_g-1}
E_g=-\frac{\tilde{t}}{2}\left(\left|\cos\frac{\phi+\theta}{2}\right|+\left|\cos\frac{\phi-\theta}{2}\right|\right),
\end{eqnarray}
whose minimum is located 
at $\phi=0$ for $\theta \in (-\pi/2+2n\pi,\pi/2+2n\pi)$ and at $\phi=\pi$ for $\theta \in (\pi/2+2n\pi,3\pi/2+2n\pi)$. At the transition points $\theta=(2n+1)\pi/2$, the topological JJ has a double degeneracy at $\phi=0$ and $\phi=\pi$. Therefore, SOC can lead to a transition between the Josephson $0$- and $\pi$-junction in this case. In Fig.~\ref{misalinged}(b), given $\theta=0, \pi/4,\pi/2, 3\pi/4, \pi$, we plot the ground state energy as a function of $\phi$ based on Eq.~(\ref{E_g-1}). 
The blue line for $\theta=\pi/2$ shows the double degeneracy at $\phi=0$ and $\phi=\pi$.

When the SOC field direction $\bm{\hat{n}}$ is perpendicular to the MF spin 
{\bf ($\beta=\pi/2$),} the MF coupling Hamiltonian depends on the azimuthal angle $\alpha$, according to Eq.~(\ref{Ht-20}).  
With $\alpha=0$,  
the  Hamiltonian reads
\begin{eqnarray*}
\tilde{H}_{t}=i \frac{\tilde{t}}{2} \gamma_{1}^a(\cos\frac{\phi}{2}\cos\frac{\theta}{2}\sigma_0-\sin\frac{\phi}{2}\sin\frac{\theta}{2}\sigma_x)\gamma_{2}^{b},
\end{eqnarray*}
whose eigenenergies  are 
\begin{eqnarray}\label{AL-4}
E&=& \pm\frac{\tilde{t}}{2}(\cos\frac{\phi}{2}\cos\frac{\theta}{2}\pm \sin\frac{\phi}{2}\sin\frac{\theta}{2}) \nonumber \\
&=&\pm \frac{\tilde{t}}{2}\cos\frac{\phi \pm \theta}{2},
\end{eqnarray} 
the same as Eq.~(\ref{AL-3}) derived for the SOC field direction parallel with the MF spin. 

For $\alpha=\pi/2$,  
the MF coupling Hamiltonian is
\begin{eqnarray}\label{Ht-22}
\tilde{H}_{t}=\frac{i\tilde{t}}{2} \cos\frac{\phi}{2} (\gamma_{1\uparrow}^a \gamma_{2\nearrow}^b+\gamma_{1\downarrow}^a \gamma_{2\swarrow}^b), \end{eqnarray}
with
\begin{eqnarray*}
\gamma_{2\nearrow}^b&=&\cos\frac{\theta}{2}\gamma_{2\uparrow}^b+\sin\frac{\theta}{2}\gamma_{2\downarrow}^b, \nonumber \\
\gamma_{2\swarrow}^b&=&\cos\frac{\theta}{2} \gamma_{2\downarrow}^b-\sin\frac{\theta}{2}\gamma_{2\uparrow}^b.
\end{eqnarray*}
The associated Andreev levels and Josephson current are
\begin{eqnarray*}
E=\pm \frac{\tilde{t}}{2}\cos\frac{\phi}{2}, \ \ I=\mp\frac{e\tilde{t}}{\hbar}\sin\frac{\phi}{2},
\end{eqnarray*} 
which behave exactly in the same manner as those in the JJ without SOC in the normal region. Moreover, the coupling strength is completely independent of the SOC.

To better understand the MF coupling Hamiltonian in the three SOC field directions, we study the $\vec{d}$-vector of MZM-induced superconducting condensates. In the normal region of a JJ, the $\vec{d}$-vector of the spin-triplet pairing will experience a torque $\lambda \hbar k_{\text f} \bm{\hat{n}} \times \vec{d}$ \cite{SC:Bergeret2013,SC:Liu2014}. For the time-reversal invariant case, at site $a$, the two MZMs $\gamma_{1\uparrow}^a,\gamma_{1\downarrow}^a$ induce the spin-triplet pairing amplitudes $\Psi_{\uparrow\uparrow}=\Psi_{\downarrow\downarrow}=1$. According to Eq.~\eqref{d-vector-2}, the corresponding $\vec{d}$-vector is $(0,i,0)$, 
which is along the $y$ axis.
When the SOC field direction $\bm{\hat{n}}$ is along the $z$ or $x$ axis, this $\vec{d}$-vector will precess in the $x-y$ or $y-z$ plane respectively with the same precession speed because of the same strength of the SOC induced torque. 
This gives similar Andreev level shift in the $\phi$ axis due to the spin-phase $\theta$ (Fig.~\ref{misalinged}(a)). When $\bm{\hat{n}}$ is along $y$ direction, the SOC-induced torque is zero, and $\vec{d}$-vector being $(0,i,0)$ will not precess so that the JJ behaves like the ones without SOC in the normal region.

\section{All electrical control and Majorana braiding}
\label{non-Abelian}

\begin{figure}[htbp]
\centering
\begin{tabular}{l}
\includegraphics[width=0.95\columnwidth]{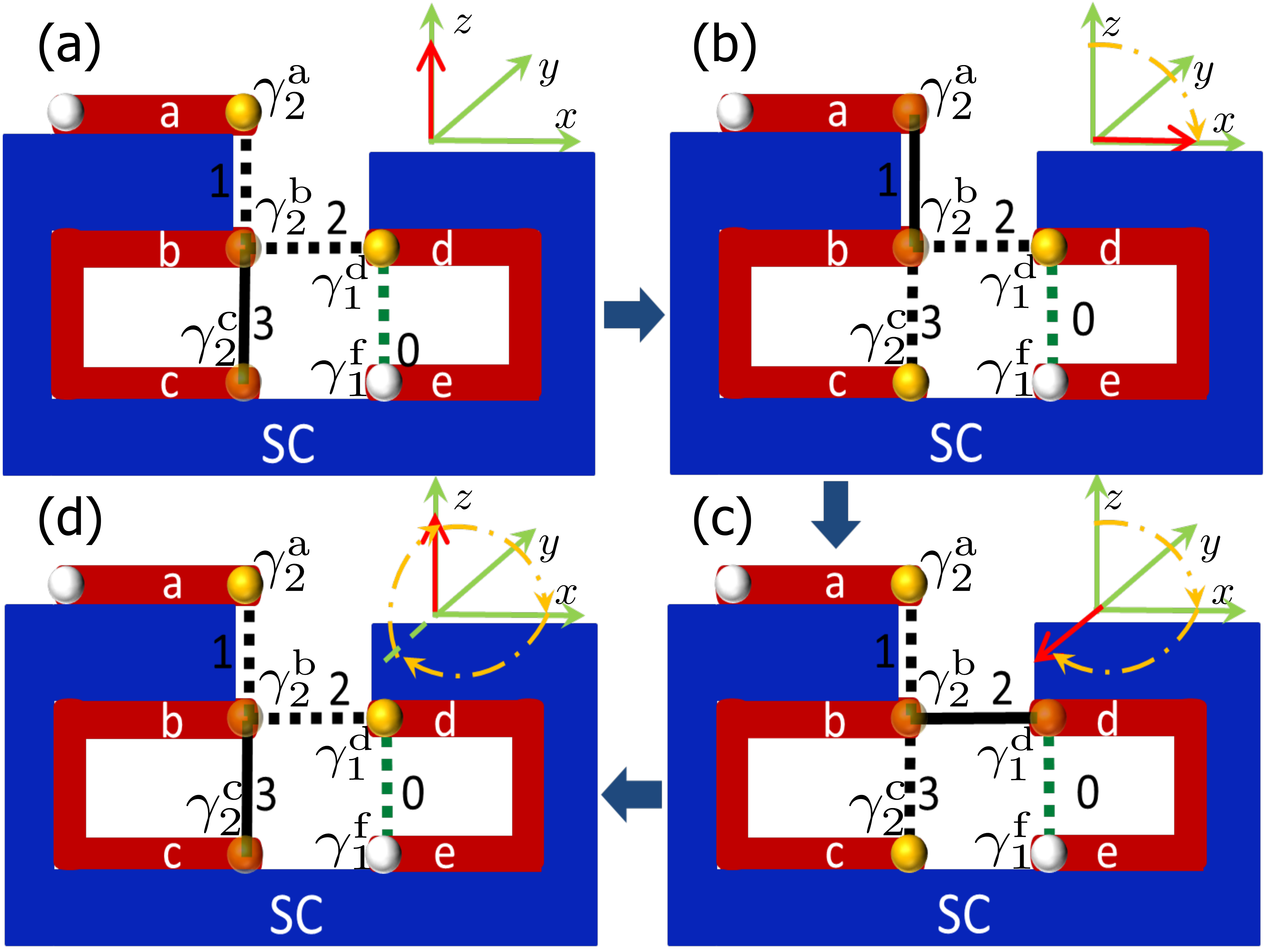}
\end{tabular}
\caption{Graphical representation of braiding MZMs and associated effective magnetic field. The three black wires labeled as 1, 2 and 3 couple the four MZMs (yellow balls). Color white for the other two MZMs is used to distinguish them from the four MZMs which are involved in the braiding process. The solid (semitransparent) balls indicate that the MZMs are uncoupled (coupled). The solid (dashed) lines represent the \lq\lq{}on\rq\rq{} and \lq\lq{}off\rq\rq{} of the coupling through the wires. The red arrows indicate the effective magnetic field direction of the braiding Hamiltonian. The yellow arrows illustrate the trajectory of the effective magnetic field during the braiding process. }
\label{braiding}
\end{figure}

\begin{figure}[htbp]
\centering
\begin{tabular}{l}
\includegraphics[width=0.8\columnwidth]{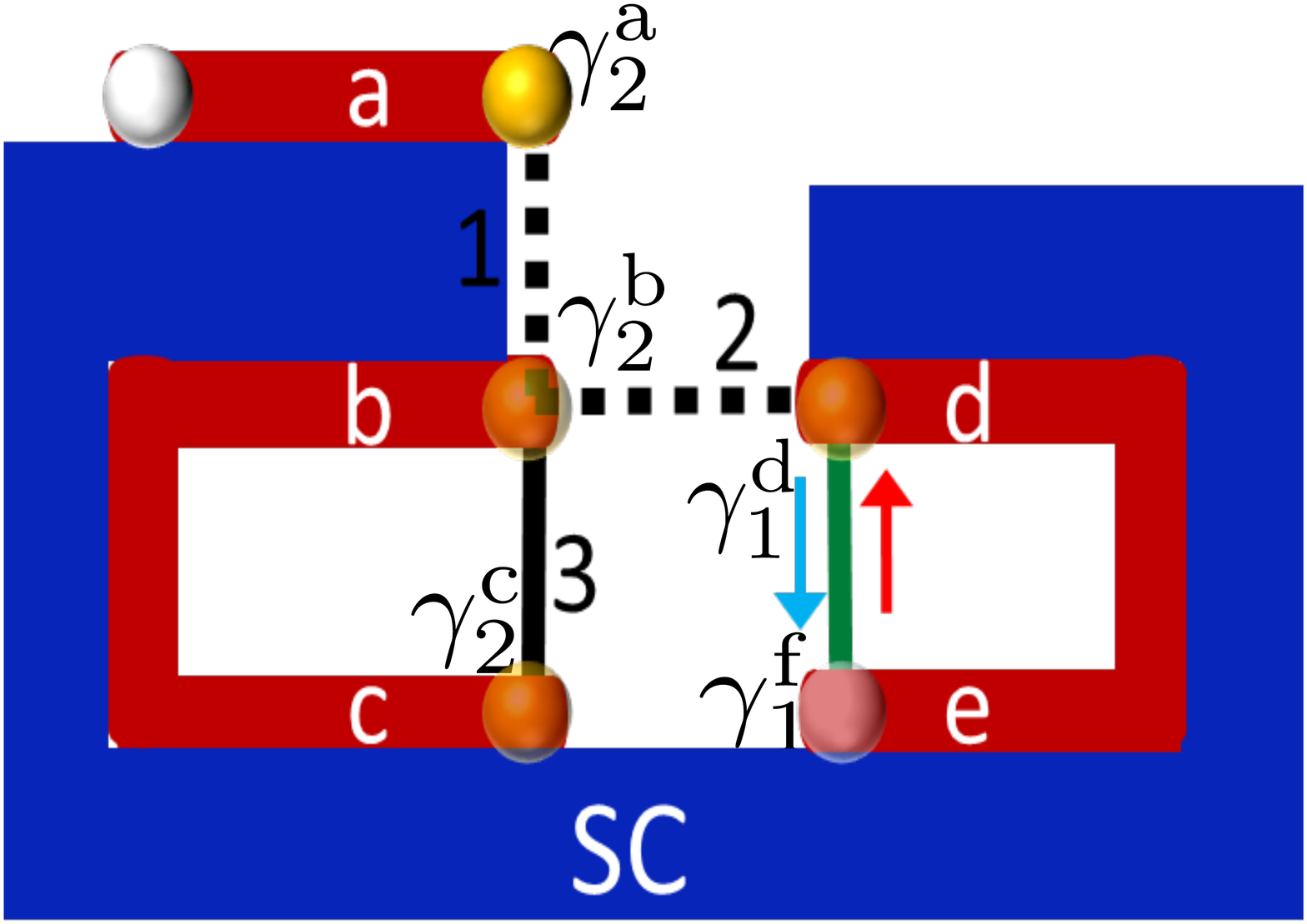}
\end{tabular}
\caption{Reading out the fermion parity by measuring the Josephson current direction. The green wire couples the MZMs $\gamma_1^{\rm d}$ and $\gamma_1^{\rm e}$ and is labeled as \lq\lq{}0\rq\rq{}. The blue and red arrows illustrate that the Josephson current changes its direction before and after braiding process due to the non-Abelian nature of MZMs.}
\label{initial}
\end{figure}

In this section, using the spin-phase physics developed in the earlier sections of this paper, we construct a semiconductor circuit to braid MZMs of time-reversal broken TSCs by gate voltage tunable MF coupling with exponential sensitivity and detect their non-Abelian statistics by measuring the SOC-driven Josephson current(Fig.~\ref{braiding}). We believe our suggested JJ-based braiding experiment to be both conceptually the most straightforward and experimentally the most practical for semiconductor Majorana nanowire systems being extensively studied in many laboratories all over the world. The basic building block of this circuit is the topological Josephson junction which connects two MZMs (Fig.~\ref{braiding}). The red wires represent semiconductor-wire based topological superconductors \cite{TSC:Sau2010,TSC:Lutchyn2010} which are coupled to the same $s$-wave superconductor so that they have the same charge phase. The three black wires couple the four MZMs (indicated by the yellow balls in Fig.~\ref{braiding}) and are used to braid $\gamma_{2}^{\rm a}$ and $\gamma_{1}^{\rm d}$. The solid (dashed) lines represent turning on (off) the MF coupling which can be exponentially accurate if the gate-tunable chemical potential of these wires is inside the semiconductor band gap as shown in the inset of Fig.~\ref{H-JJ}(b). The green line, coupling $\gamma_{1}^{\rm d}$ with $\gamma_{1}^{\rm e}$ (Fig.~\ref{braiding}), is turned on before and after braiding to detect the non-Abelian statistics as being illustrated below (Fig.~\ref{initial}). 

During the braiding process (Fig.~\ref{braiding}(a,b,c,d)) finite SOC is necessary to couple $\gamma_2^{\rm b}$ with $\gamma_2^{\rm a}$ and $\gamma_2^{\rm c}$ (wires 1 and 3 in Fig.~\ref{braiding}) which form topological $\pi$ JJs with the associated Hamiltonian
\begin{eqnarray}\label{Ham-couple-1}
H_{1(3)}=i \tilde{t}_{1(3)} \gamma_{2}^{\rm a(c)}\gamma_{2}^{\rm b}\sin\frac{\theta}{2}.
\end{eqnarray}  
Here $\tilde{t}_{1(3)}$ is the tunneling amplitude which is exponentially sensitive to the wire chemical potential inside the semiconductor band gap (inset of Fig.~\ref{H-JJ}(b)) and $\theta$ is the spin-phase defined in Eq. \eqref{phase-1}. For simplicity, $\theta$ is assumed to be the same in these two wires without loss of generality. The Zeeman term is parallel to the SOC direction in wires 0, 1 and 3, and thereby only effectively shifts the chemical potential of electrons in the same spin channel without affecting the spin-phase across these wires. The coupling between $\gamma_{2}^{\rm b}$ and $\gamma_1^{\rm d}$ forms a topological $0$-junction and SOC is not necessary for having a finite MZM coupling strength in wire 2 (Fig.~\ref{braiding}). Accordingly, the Hamiltonian of wire 2 has the form
\begin{eqnarray}\label{Ham-couple-2}
H_{2}=i \tilde{t}_2 \gamma_{2}^{\rm b}\gamma_{1}^{\rm d}.
\end{eqnarray}

During the braiding process, the states corresponding to the four MZMs $\gamma_2^{\rm a}$, $\gamma_2^{\rm b}$, $\gamma_2^{\rm c}$ and $\gamma_1^{d}$ (yellow balls in Fig.~\ref{braiding}) can be written in the Fock basis as \begin{eqnarray*}
|00\rangle, |11\rangle=c_{2}^{\dagger}c_1^{\dagger}|00\rangle, |01\rangle=c^{\dagger}_2|00\rangle, |10\rangle=c^{\dagger}_1 | 00 \rangle  \end{eqnarray*} with the occupation numbers of the two fermionic operators $c_1=(\gamma_2^{\rm a}-i\gamma_1^{\rm d})/2$ and $c_2=(\gamma_2^{\rm c}-i\gamma_2^{\rm b})/2$. Here $|0\rangle$ and $|1\rangle$ correspond to the $+$ and $-$ of the fermion parity $1-2c^{\dagger}c$, respectively. The total fermion parity of these four MZMs is $\mathcal{P}(t)=\langle t | i\gamma_2^{\rm a}\gamma_1^{\rm d}i\gamma_2^{\rm b}\gamma_2^{\rm c} | t \rangle$ with $\langle t| \cdots | t \rangle$ the average of the state at time $t$. The MF coupling Hamiltonians through the three black wires (Fig.~\ref{braiding}) in the Fock basis have the form
\begin{widetext}
\begin{eqnarray}
\label{Ham-couple-3}
H_{br}=H_1+H_2+H_3=\left(\begin{array}{cc} (\tilde{t}_2  s_y-\tilde{t}_1 s_x)\sin\frac{\theta}{2}-\tilde{t}_3 s_z & 0 \\ 0 &(\tilde{t}_2 s_y-\tilde{t}_1 s_x)\sin\frac{\theta}{2}+\tilde{t}_3 s_z \end{array}\right), 
\end{eqnarray} 
\end{widetext} where $s_{x,y,z}$ are the three Pauli matrices acting on the subspace spanned by ($|00\rangle$, $|11\rangle$) with even total fermion parity or ($|01\rangle$, $|10\rangle$) with odd total fermion parity. The Hamiltonian $H_{br}$ in Eq.~\eqref{Ham-couple-3} is block diagonal which indicates the conservation of the total fermion parity during the braiding process. In each block, the Hamiltonian is exactly the same as that of a spin-$\frac{1}{2}$ particle in a magnetic field (red arrows in Fig.~\ref{braiding}). For simplicity, we assume $\tilde{t}_{1,2,3}$ are positive and $\theta \in (0,\pi)$. The red arrows in Fig.~\ref{braiding} illustrate the effective magnetic field in the block of the even total fermion parity with basis $|00\rangle$ and $|11\rangle$. The coupling of the four MZMs through the Hamiltonian Eq.~\eqref{Ham-couple-3} is equivalent to the tri-junction discussed extensively in the semiconductor Majorana circuit literature \cite{TSC:Sau2011,TSC:Heck2012,TSC:Hyart2013}. Accordingly, the braiding operation can be realized by turning on and off the coupling in wires 1, 2 and 3 sequentially as shown in Fig.~\ref{braiding}. 

In the beginning, only the coupling in wire 3 is on so that the effective magnetic field is along $z$ direction (Fig.~\ref{braiding}(a)). We first turn off the coupling in wire 3 and in the meantime turn on the coupling in wire 1 so that the MZM $\gamma_2^{\rm a}$ is transported to $\gamma_{2}^{\rm c}$ and the effective magnetic field is along $x$ direction (Fig.~\ref{braiding}(b)). Then we turn off the coupling in wire 1 and in the meantime turn on the coupling in wire 2. The MZM $\gamma_1^{\rm d}$ is then transported to $\gamma_2^{\rm a}$ and the effective magnetic field is along $-y$ direction (Fig.~\ref{braiding}(c)). At last, we turn off the coupling in wire 2 and at the same time turn on the coupling in wire 3 so that the MZM is transported from $\gamma_2^{\rm c}$ to $\gamma_1^{\rm d}$ and the effective magnetic field comes back to its original direction (Fig.~\ref{braiding}(d)). During the braiding operation, the effective field encloses a solid angle $\pi/2$ as illustrated in Fig.~\ref{braiding}. Consequently, the evolution operator takes the form \cite{Sakurai2011}
\begin{eqnarray}\label{evo}
U=\left(\begin{array}{cc}\exp(-i\frac{\pi}{4}s_z) & 0 \\ 0 & \exp(-i\frac{\pi}{4} s_z) \end{array}\right),
\end{eqnarray}
and the MZMs $\gamma_2^{\rm a}$ and $\gamma_1^{\rm d}$ in the Heisenberg representation are transformed as
\begin{eqnarray}\label{evo-MZM}
\gamma_2^{\rm a}(T)&=&U\gamma_2^{\rm a} U^{\dagger}=-\gamma_1^{\rm d}, \nonumber \\ \gamma_1^{\rm d}(T)&=&U\gamma_1^{\rm d} U^{\dagger}=\gamma_2^{\rm a},
\end{eqnarray}
with $T$ the braiding time. 

To detect the non-Abelian braiding statistics, we connect $\gamma_{1}^{\rm d}$ with $\gamma_1^{\rm e}$ through the wire 0 (green wire in Fig.~\ref{initial}). According to the discussion in Sec.~\ref{TRB-TJJ}, given a finite spin-phase, the fermion parity $\mathcal{P}_0$ is locked to the sign of the fractional Josephson current through the wire 0. As the braiding is operated in the absence of charge phase, we expect the normal Cooper pair tunneling with $2\pi$ periodicity will not contribute to the Josephson current. Thus, the initial fermion parity in wire 0, $\mathcal{P}_0(t=0)=\langle 0 | i\gamma_{1}^a \gamma_{1}^b | 0 \rangle$, can be detected by measuring the Josephson current through the wire 0. We then turn off the coupling in wire 0 and start to braid the MZMs $\gamma_2^{\rm a}$ and $\gamma_1^{\rm d}$ following the procedure in Fig.~\ref{braiding}. After braiding the MZMs twice, according to Eq.~\eqref{evo-MZM}, the evolution of MZMs in Heisenberg representation satisfies
\begin{eqnarray*}
\gamma_{1}^{\rm d}(2T)=U^2\gamma_{1}^{\rm d}(U^{\dagger})^2=-\gamma_{1}^{\rm d}.
\end{eqnarray*}
At the same time, as the coupling in wire 0 is turned off during the process, the MZM $\gamma_1^b$ commutes with the Hamiltonian $H_{br}$ so that $\gamma_1^b(2T)=\gamma_1^b$.
The fermion parity in wire 0 in this case is
\begin{eqnarray*}
\mathcal{P}_0(2T) &=&\langle 2T | i\gamma_1^{\rm d} \gamma_1^{\rm e} | 2T \rangle = \langle 0 | (U^{\dagger})^2 i\gamma_1^{\rm d} \gamma_1^{\rm e} U^2 |0\rangle \nonumber \\
&=& \langle 0 | i\gamma_{1}^{\rm d}(2T) \gamma_{1}^{\rm e}(2T) |0\rangle=-\langle 0 | i\gamma_1^{\rm d} \gamma_1^{\rm e} |0\rangle \nonumber \\
&=& -\mathcal{P}_0(0),
\end{eqnarray*}
which is opposite to its initial value.
Consequently, when we turn on the coupling in wire 0 after braiding twice, the Josephson current direction should also be opposite to that before braiding. Therefore, the non-Abelian braiding statistics can be directly probed simply by measuring the spin-phase driven Josephson current direction in wire 0 before and after braiding.

{\it Experimental feasibility:} We note here that the U-junction is the building block of our proposed braiding devices. Since the single-crystaline InSb nanowire networks \cite{TSC:Car2014} has been fabricated and semiconductor nanowires with SOC coupling, proximity-induced superconducting gap, and Zeeman spin
splitting have been realized experimentally \cite{TSC:Mourik2012,TSC:Deng2012,TSC:Rokhinson2012,TSC:Xu2014}, there should be no significant technical obstacle to realize the proposed devices in Fig.~\ref{braiding} and Fig.~\ref{initial}. For the braiding process, our proposal only needs to tune the spin-orbit coupling through the gate voltages inside the normal nanowire region (black and green wires in Fig.~\ref{braiding} and Fig.~\ref{initial} rather than the region of Majorana wires (red wires) in proximity to the superconductors. Thus the proposed gate-control of spin-orbit coupling in our work can be achieved by standard technologies of the semiconductor spintronics \cite{Weber:2007_a,Koralek:2009_a,Wunderlich2010,Yang2012,Walser2012,Dettwiler2014}. The measured Josephson current, according to our calculation shown in Fig.~\ref{cvm}, is about $0.07 (2e\Delta/\hbar)$ with $\Delta$ the proximity induced superconductors gap in the topological superconducting nanowires (red wires in Fig.~\ref{braiding} and Fig.~\ref{initial}). By taking the proximity induced hard gap in InAs as 0.2 meV \cite{TSC:ChangW2015}, the maximal SOC current in our model is about 2nA within the experimentally detectable regime. Thus our proposal has the advantage of experimental feasibility. 

\section{Discussion and conclusion} 

In conclusion, we have established the emergence of {\it spin-phase} for MFs and demonstrated that it can be tuned by SOC in the normal region of the topological JJ.  we theoretically proposed several applications in Majorana spintronics. This spin-phase has an important topological origin in the sense that it arises from an intrinsic Majorana induced $\pi$-phase difference between opposite ends of a spinful one-dimensional TSC, which we have shown to exhibit crucial spin dependence. The spin-phase based Majorana spin physics is not only conceptually novel, but also implies innovative applications in making robust gate voltage tunable fractional $\pi$- and $\phi_0$-junctions. We thus provide an all-electrically controlled semiconductor circuit to braid MZMs and to read out the topological information encoded by their non-Abelian statistical features without tuning any external magnetic flux, which opens the way to electrically controlled Majorana spintronics studies. 

We acknowledge helpful discussions with Jay D. Sau, P. M. R. Brydon, Dong E. Liu and Zheng-Xin Liu. This work is supported by LPS-MPO-CMTC and Microsoft Q. X. J. L is supported by National Natural Science Foundation of China (Grant No. 11574008), and by the Thousand-Young-Talent Program of China.

\appendix

\begin{widetext}

\section{Spin-triplet Cooper pairs}\label{spin-triplet-pair}

Considering a Cooper pair composed of two electrons, the wave function can be heuristically written as 
\begin{eqnarray}\label{spin1-1}
\Psi=\psi_{\uparrow\uparrow}|\uparrow\uparrow\rangle+\psi_{\uparrow\downarrow}|\uparrow\downarrow\rangle+\psi_{\downarrow\uparrow}\downarrow\uparrow\rangle+\psi_{\downarrow\downarrow}|\downarrow\downarrow\rangle,
\end{eqnarray}
where $\psi_{\uparrow\uparrow(\uparrow\downarrow,\downarrow\uparrow,\downarrow\downarrow)}$ are the coefficients of the corresponding spin states. 
However, 
Eq.~(\ref{spin1-1}) is not convenient or precise in describing superconductors
lacking spin rotation symmetry. 
To get a more general and concise description of the spin-triplet Cooper pairs, 
we introduce a four-component spinor field $C(\hat{x}_+)=[c_{\uparrow}(\hat{x}_+),c_{\downarrow}(\hat{x}_+),-c_{\downarrow}^{\dagger}(\hat{x}_-),c_{\uparrow}^{\dagger}(\hat{x}_-)]^{\rm T}$, where $\hat{x}_{\pm}=(\pm\epsilon,\bm r)$ with $\epsilon$ and $\bm r$ the energy and spatial coordinates respectively. Then the density matrix in the second quantization language can be defined as
\begin{eqnarray*}
D_{ij}(\hat{x}_+)&=&-iG^{<}_{ij}(\hat{x}_+)=\langle C^{\dagger}_j(\hat{x}_+)\otimes C_i(\hat{x}_+)\rangle,
\end{eqnarray*}
where
$G^{<}$ is the lessor Green's function \cite{Rammer:2007_a}.
The anomalous density matrix (with elements taken from the up-right two-by-two block of the four-by-four density matrix $D(\hat{x}_+)$) that corresponds to the pairing amplitudes can be written as
\begin{eqnarray}\label{spin1-3}
f(\hat{x}_+)=\left(\begin{array}{cc} -\langle c_{\downarrow}(\hat{x}_-) c_{\uparrow}(\hat{x}_+)\rangle & \langle c_{\uparrow}(\hat{x}_-) c_{\uparrow}(\hat{x}_+)\rangle \\ -\langle c_{\downarrow}(\hat{x}_-) c_{\downarrow}(\hat{x}_+)\rangle & \langle c_{\uparrow}(\hat{x}_-) c_{\downarrow}(\hat{x}_+)\rangle \end{array}\right) 
=\left(\begin{array}{cc} -\psi_{\downarrow\uparrow} & \psi_{\uparrow\uparrow} \\ -\psi_{\downarrow\downarrow} &\psi_{\uparrow\downarrow} \end{array}\right).
\end{eqnarray}
Here $\langle \dots \rangle$ represents the thermal average. When electrons are coupled to a magnetic flux, as spin-up and spin-down electrons have the same charge $e$, they will acquire the same charge phase $\phi(\bm r)/2$ as
\begin{eqnarray*}
\left(\begin{array}{c} c_{\uparrow}(\bm r) \\ c_{\downarrow}(\bm r) \end{array}\right) \longrightarrow \left(\begin{array}{c} c_{\uparrow}(\bm r) \\ c_{\downarrow}(\bm r) \end{array}\right) e^{i\phi(\bm r)/2}.
\end{eqnarray*}
According to Eq.~\eqref{spin1-3}, the anomalous density matrix is transformed as
\begin{eqnarray}\label{spin1-4}
\left(\begin{array}{cc} -\psi_{\downarrow \uparrow} & \psi_{\uparrow\uparrow} \\ -\psi_{\downarrow\downarrow} &\psi_{\uparrow\downarrow} \end{array}\right)
\longrightarrow \left(\begin{array}{cc} -\psi_{\downarrow\uparrow} & \psi_{\uparrow\uparrow} \\ -\psi_{\downarrow\downarrow} &\psi_{\uparrow\downarrow} \end{array}\right) e^{i\phi(\bm r)},
\end{eqnarray}
which indicates that all pairing states must have the same charge phase.

By using the anomalous density matrix, we can define a complex vector, 
\begin{eqnarray}\label{d-vector-1}
\vec{d}=\rm{Tr}\left[\frac{1}{2}\bm{\sigma}f\right],
\end{eqnarray}
with $\bm{\sigma}=(\sigma_x,\sigma_y,\sigma_z)$ the Pauli matrices in spin space.
From Eq.~(\ref{d-vector-1}), 
we have 
\begin{eqnarray}\label{d-vector-2}
d_x=\frac{\psi_{\uparrow\uparrow}-\psi_{\downarrow\downarrow}}{2}, d_y=i\frac{\psi_{\uparrow\uparrow}+\psi_{\downarrow\downarrow}}{2}, d_z=-\frac{\psi_{\uparrow\downarrow}-\psi_{\downarrow\uparrow}}{2}.
\end{eqnarray}
With the $\vec{d}$-vector representation, 
it is instructive to
consider a spin-triplet state with coefficients $\psi_{\uparrow\uparrow}=e^{-i\theta},\psi_{\uparrow\downarrow}=\psi_{\downarrow\uparrow}=0,\psi_{\downarrow\downarrow}=e^{i\theta}$. The corresponding anomalous part of the density matrix is
\begin{eqnarray}\label{D-2}
f=\left(\begin{array}{cc} 0&e^{-i\theta} \\ -e^{i\theta} & 0 \end{array}\right)=i(\cos\theta\sigma_y-\sin\theta\sigma_x),
\end{eqnarray}
with $(d_x,d_y,d_z)^{\text T}=i(-\sin\theta,\cos\theta,0)^{\text T}$. We find that the relative phase between the state $|\uparrow\uparrow\rangle$ and $|\downarrow\downarrow\rangle$ corresponds to the azimuthal angle of the $\vec{d}$-vector in the $x$-$y$ plane. It is worthwhile to note that this state has no spin polarization although it is 
a spin-triplet state. This type of spin-triplet state is called unitary state \cite{SC:Leggett1975}. 
For any unitary spin-triplet state, its
$\vec{d}$-vector is always  real up to an overall U(1) phase~\cite{SC:Leggett1975}
and therefore it satisfies $|\vec{d}\cdot\vec{d}|=|d_x|^2+|d_y|^2+|d_z|^2$. 
There is yet another type of spin-triple state whose pairing is restricted in one spin channel, 
and such states are
fully spin polarized. For a state of this type, we can always choose a spin basis with
$\psi_{\uparrow\uparrow}=1$ and $\psi_{\uparrow\downarrow}=\psi_{\downarrow\uparrow}=\psi_{\downarrow\downarrow}=0$. 
The corresponding anomalous density matrix reads
\begin{eqnarray}\label{D-3}
f=\left(\begin{array}{cc} 0&1\\ 0 & 0 \end{array}\right)=\frac{1}{2}\sigma_x+\frac{i}{2}\sigma_y,
\end{eqnarray} 
and we have
$(d_x,d_y,d_z)=(1/2,i/2,0)$. The $\vec{d}$-vector of this state satisfies the condition $\vec{d}\cdot \vec{d}=0$.

\section{$0$-$\pi$ topological JJ transition beyond the strong Zeeman splitting limit}\label{0-pi}

In this section, we consider the transformation of Cooper pairs under the mirror operation $\mathcal{M}_z$. Cooper pairs can be described by a block off-diagonal density matrix,
\begin{eqnarray}\label{appen-density-1}
D^{\rm off}(\bm r)=\left(\begin{array}{cc} 0 & f(\bm r) \\ \bar{f}((\bm r) & 0 \end{array}\right)=\left(\begin{array}{cc} 0 & (d_0((\bm r)\sigma_0+d_i((\bm r) \sigma_i) i\sigma_y \\  -i\sigma_y(d_0((\bm r)^*\sigma_0+d_i^*((\bm r) \sigma_i) & 0 \end{array}\right),
\end{eqnarray}
with $\bm r =(x,y,z)$ and $d_0((\bm r)$ and $d_{i=x,y,z}(\bm r)$ the amplitudes of $s$-wave spin-singlet Cooper pairs and spin-triplet triplet ones with $\vec{d}$-vector along $x$, $y$ and $z$ direction respectively. Here, we assume that the spin-triplet pairs are induced by MZMs and are thereby $s$-wave \cite{TSC:Asano2013,TSC:Liu2015}. The mirror operator $\mathcal{M}_z$ in electron-hole and spin spaces is 
\begin{eqnarray*}
\mathcal{M}_z=\left(\begin{array}{cc} i\sigma_z & 0 \\ 0 & -i\sigma_z \end{array}\right).
\end{eqnarray*}
For a mirror ($\mathcal{M}_z$) symmetric system we have
\begin{eqnarray}\label{appen-density-2}
\mathcal{M}_z D^{\rm off}(\bm r) \mathcal{M}_z^{-1} =\left( \begin{array}{cc} 0 & i\sigma_z f(x,y,z) i\sigma_z \\ i\sigma_z \bar{f}(x,y,z) i\sigma_z & 0 \end{array}\right)=\left(\begin{array}{cc} 0 & f(x,y,-z) \\ \bar{f}(x,y,-z) & 0 \end{array}\right),
\end{eqnarray}
with 
\begin{eqnarray*} 
i\sigma_z f(\bm r) i\sigma_z &=& (d_0(\bm r)\sigma_0+d_z(\bm r) \sigma_z-d_x(\bm r)\sigma_x-d_y(\bm r)\sigma_y) i\sigma_y \nonumber \\
&=&f(x,y,-z)=(d_0(x,y,-z)\sigma_0+d_z(x,y,-z) \sigma_z+d_x(x,y,-z)\sigma_x+d_y(x,y,-z)\sigma_y) i\sigma_y.
\end{eqnarray*}
As the magnetic flux does not distinguish spin-up or spin-down, 
all of the four Cooper pairs should have the same flux-induced charge phase. 
However according to Eq.~\eqref{appen-density-2}, only those spin-triplet pairs, with $\vec{d}$-vector along the $x$ or $y$ direction,
are opposite in sign at $z$ and $-z$. In other words, they have a $\pi$-phase difference. Besides, the $s$-wave spin-singlet bulk superconducting gap in the semiconductor nanowire models (Eq.~\eqref{Ham-SOC-1} and Eq.~\eqref{gap-iron}) is uniform in the entire wire.
We therefore conclude here that this $\pi$-phase difference for the $d_x$ and $d_y$ spin-triplet Cooper pairs is not a charge phase but a spin-phase.

We have demonstrated topological $0$ and $\pi$ JJs in the strong Zeeman splitting limit of a 1D BDI class topological superconductor model in Sec.~\ref{0-pi-sec}. Actually the conclusion of topological $0$ and $\pi$ JJs for the setups in Figs.~\ref{D-current-phase}(a) and \ref{D-current-phase}(b) is valid for any multi-channel D class topological superconductor model as long as it has mirror symmetry with mirror plane perpendicular to the wire direction. 

We first consider the 1D BDI class topological superconductor model beyond the strong Zeeman splitting limit. The tight-binding model described by Eq.~\eqref{Ham-SOC-1} in the continuous limit takes the form
\begin{eqnarray}\label{appen-Ham-1}
H=\left[(-\frac{\hbar^2}{2m} \partial_z^2-\mu) \sigma_0+M_z \sigma_z+\lambda i\partial_z \sigma_y \right] \tau_z-\Delta \sigma_y\tau_y.
\end{eqnarray}
For the realistic semiconductor nanowire, the chemical potential $\mu \approx 0$ and the SOC energy $\lambda k_f$ is much smaller than the superconducting gap $\Delta$ and Zeeman energy $M_z$ \cite{TSC:Mourik2012,TSC:Sarma2015}. In this case, the right MZM of Eq.~\eqref{appen-Ham-1} is \cite{TSC:DasSarma2012,TSC:He2014}
\begin{eqnarray}\label{appen-wf-1}
\gamma_{\rm 1}(z)&=&\left(\begin{array}{cc} \hat{c} & \hat{c}^{\dagger}  \end{array}\right)\left(\begin{array}{c} u_{1}(z) \\ u_{1}(z) \end{array}\right),
\end{eqnarray}
where \begin{eqnarray*}
\hat{c}= \left( \begin{array} {c} c_{\uparrow} \\ c_{\downarrow} \end{array} \right), \ \  u_{1}(z)=u_{1}^{*}(z) = e^{wz}\left(\begin{array}{c} \Delta+w\lambda \\ M_z-\mu-w^2 \end{array} \right)+ j e^{vz}\left(\begin{array}{c} \Delta+v\lambda \\ M_z-\mu-v^2 \end{array} \right)+ j^* e^{v^* z}\left(\begin{array}{c} \Delta+v^*\lambda \\ M_z-\mu-v^{*2} \end{array} \right),
\end{eqnarray*}
$v \approx i k_{\rm f, eff}+\delta$ and $w \approx k_{\rm f,eff}+\delta$ with $k_{\rm f, eff} \approx \sqrt{2m}(M_z^2-\Delta^2)^{1/4}$ and $\delta \approx \lambda \Delta m/\sqrt{V_z^2-\Delta^2}$. To satisfy the boundary condition $\gamma_1(z=0)=0$, we have $j=-1/2-i/2$. 

As $e^{wz}$ decays much faster than $e^{vz}$ or $e^{v^*z}$, by neglecting the term containing $e^{wz}$ in Eq.~\eqref{appen-wf-1}, the MZM wave function inside the TSC can be simplified as 
\begin{eqnarray}\label{appen-wf-2}
u_{1}(z)=\cos(k_{\rm f,eff}z) e^{\delta z} \left(\begin{array}{c} \Delta \\ M_z+\sqrt{M_z^2-\Delta^2} \end{array} \right)=s(z) \left(\begin{array}{c} \cos(\frac{\eta}{2}) \\ \sin(\frac{\eta}{2}) \end{array} \right),
\end{eqnarray}
where its spin direction only depends on $\Delta$ and $M_z$ with $\tan\eta=\Delta/(M_z+\sqrt{M_z^2-\Delta^2})$. Here, because SOC energy is much smaller than the Zeeman energy and superconducting gap, we also neglect the terms containing SOC strength $\lambda$.  As the system is invariant under the mirror operation $\mathcal{M}_z$, the left MZM $\gamma_2$ takes the form
\begin{eqnarray*}
\gamma_{2}(z)&=& \left(\begin{array}{cc} \hat{c} & \hat{c}^{\dagger}  \end{array}\right)\left(\begin{array}{c} i u_{2}(z) \\ -i u_{2}(z) \end{array}\right), \nonumber \\
u_{2}(z)&=&i\sigma_z u_{1}(-z)= s(-z) \left(\begin{array}{c} \cos(\frac{\eta}{2})\\ -\sin(\frac{\eta}{2}) \end{array} \right).
 \end{eqnarray*} 
 
For the topological JJ in Fig.~\ref{D-current-phase}(a), the Hamiltonian in Eq.~\eqref{Ht-2-1} can be rewritten in the spin quantization basis of $\gamma_{1}^a$ and $\gamma_{2}^b$ as
 \begin{eqnarray}\label{apd-Ham-1}
 \tilde{H}(t)&=&\tilde{t}  \left( \begin{array}{c} \cos(\frac{\eta}{2}) c_{a,\uparrow}^{\dagger} \\ \sin(\frac{\eta}{2}) c_{a,\downarrow}^{\dagger} \\ \cos(\frac{\eta}{2}) c_{a,\uparrow} \\ \sin(\frac{\eta}{2}) c_{a,\downarrow}  \end{array} \right)^{\rm T}  \left(\begin{array}{cc} e^{\frac{i}{2}( \phi \sigma_0 )} & 0 \\ 0 & -e^{-\frac{i}{2}( \phi \sigma_0 )} \end{array}\right) \left( \begin{array}{c} i\cos(\frac{\eta}{2}) c_{b,\uparrow} \\ -i\sin(\frac{\eta}{2}) c_{b,\downarrow} \\ i\cos(\frac{\eta}{2}) c_{b,\uparrow} \\ -i\sin(\frac{\eta}{2}) c_{b,\downarrow} \end{array} \right) \nonumber \\
&=& \tilde{t}  \left( \begin{array}{c} \cos(\frac{\eta}{2}) c_{a,\uparrow} \\ \sin(\frac{\eta}{2}) c_{a,\downarrow} \\ \cos(\frac{\eta}{2}) c_{a,\uparrow}^{\dagger} \\ \sin(\frac{\eta}{2}) c_{a,\downarrow}^{\dagger} \ \end{array} \right)^{\rm T} \hat{R}_1^{\dagger} \hat{R_1}  \left(\begin{array}{cc} e^{\frac{i}{2}( \phi \sigma_0 )} & 0 \\ 0 & -e^{-\frac{i}{2}( \phi \sigma_0 )} \end{array}\right) \hat{R}_2^{\dagger} \hat{R}_2 \left( \begin{array}{c} i\cos(\frac{\eta}{2}) c_{b,\uparrow} \\ -i\sin(\frac{\eta}{2}) c_{b,\downarrow} \\ -i\cos(\frac{\eta}{2}) c_{b,\uparrow} \\ i\sin(\frac{\eta}{2}) c_{b,\downarrow} \end{array} \right) \nonumber \\
&=&\tilde{t} \left( \begin{array}{c} c_{a,\nearrow}^{\dagger} \\ 0 \\ c_{a,\nearrow} \\ 0 \end{array}\right)^{\rm T}  \left(\begin{array}{cc} e^{\frac{i}{2}( \phi \sigma_0+2\eta\sigma_y )} & 0 \\ 0 & -e^{-\frac{i}{2}( \phi \sigma_0-2\eta\sigma_y )} \end{array}\right) \left( \begin{array}{c} i c_{b,\nwarrow} \\ 0 \\ -i c_{b,\nwarrow}^{\dagger} \\ 0  \end{array}\right) \nonumber \\
&=& \tilde{t} \left( \begin{array}{c} c_{\nearrow}^{\dagger} \\ 0 \\ c_{\nearrow} \\ 0 \end{array}\right)^{\rm T} U^{\dagger} U  \left(\begin{array}{cc} e^{\frac{i}{2}( \phi \sigma_0+2\eta\sigma_y )} & 0 \\ 0 & -e^{-\frac{i}{2}( \phi \sigma_0-2\eta\sigma_y )} \end{array}\right) U^{\dagger} U \left( \begin{array}{c} i c_{b,\nwarrow} \\ 0 \\ -i c_{b,\nwarrow}^{\dagger} \\ 0  \end{array}\right) \nonumber \\
&=& \tilde{t}\left( \begin{array}{c} \gamma_{1,\nearrow}^{a} \\ 0 \\ 0 \\ 0 \end{array}\right)^{\rm T} \left( \begin{array}{cc} e^{i\eta \sigma_y} i\sin\frac{\phi_0}{2} & i e^{i\eta \sigma_y}  \cos \frac{\phi_0}{2} \\ -i e^{i\eta \sigma_y}  \cos \frac{\phi_0}{2} & i e^{i\eta \sigma_y} \sin \frac{\phi_0}{2} \end{array} \right) \left( \begin{array}{c} 0 \\ 0 \\ \gamma_{2,\nwarrow}^{b} \\ 0 \end{array}\right)=i\tilde{t} \cos\eta \cos\frac{\phi_0}{2}\gamma_{1\nearrow}^a \gamma_{2\nwarrow}^b,
\end{eqnarray}
where
\begin{eqnarray*}
\hat{R}_1=\exp(\frac{i}{2}\eta \sigma_y) \tau_0, \ \ \hat{R}_2=\exp(-\frac{i}{2} \eta\sigma_y) \tau_0
\end{eqnarray*}
rotate the spin quantization axis of $\gamma_1$ and $\gamma_2$ to $z$ axis respectively and 
\begin{eqnarray*}
&&\gamma_{1\nearrow}=(\cos(\eta/2)c_{\uparrow}+\sin(\eta/2) c_{\downarrow})+(\cos(\eta/2)c_{\uparrow}^{\dagger}+\sin(\eta/2) c_{\downarrow}^{\dagger}), \\ &&\gamma_{2\nwarrow}=-i(\cos(\eta/2)c_{\uparrow}-\sin(\eta/2) c_{\downarrow})+i(\cos(\eta/2)c_{\uparrow}-\sin(\eta/2) c_{\downarrow}).
 \end{eqnarray*}

For the multi-channel semiconductor wire, there is an additional SOC hopping term $\lambda \hat{p}_y \sigma_z$ which breaks the complex symmetry but respects the mirror symmetry with the mirror plane perpendicular to the wire direction. Thus the spin wave function of MZM $\gamma_1$ and $\gamma_2$ can be generally written as $(\cos(\eta/2),\sin(\eta) e^{-i\zeta})^{\rm T}$ and $(i\cos(\eta/2),-i\sin(\eta) e^{-i\zeta})^{\rm T}$ respectively with $\zeta$ the azimuthal angle of $\gamma_1$ spin polarization. Correspondingly, the rotation operators $\hat{R}_1$ and $\hat{R}_2$ are generalized to $\hat{R}_{1(2)}=\exp(\pm i \vec{n} \cdot \vec{\sigma})$ with $\vec{n}=-\sin(\zeta)\vec{e}_x+\cos(\zeta) \vec{e}_y$. By inserting the new spin wave functions and rotation operators into Eq.~\eqref{apd-Ham-1}, we find that the coupling Hamiltonian form is not affected by the azimuthal angle $\zeta$ and is thus valid for multi-channel D class TSCs which are invariant under the mirror operation $\mathcal{M}_z$.

It is noted that effective coupling Hamiltonian (Eq.~\eqref{apd-Ham-1}) is always a topological Josephson 0-junction and $\cos\eta$ is the inner product of the spin wave functions of $\gamma_{1\nearrow}^a$ and $\gamma_{2\nwarrow}^b$. When $\eta=0$, Eq.~\eqref{apd-Ham-1} returns to the MZM coupling Hamiltonian for the strong Zeeman splitting limit. For the topological JJ in Fig.~\ref{D-current-phase}(b), the two MZMs $\gamma_{1}^{a}$ and $\gamma_1^b$ have the same spin polarization so that their coupling Hamiltonian is always 
\begin{eqnarray*}
H_{t}=i\sin\frac{\phi_0}{2}\gamma_{1\nearrow}^a \gamma_{1\nearrow}^b,
\end{eqnarray*}
which corresponds to a topological Josephson $\pi$-junction. 

\end{widetext}

%

\end{document}